\documentclass[cup6b]{cupbook}

\usepackage{graphicx}
\usepackage{natbib}

\def\etal{{et\,al.}}
\def\msun{M$_{\odot}$}

\def\degs{\ifmmode ^{\circ}\else$^{\circ}$\fi}
\def\amin{\ifmmode ^{\prime}\else$^{\prime}$\fi}
\def\asec{\ifmmode ^{\prime\prime}\else$^{\prime\prime}$\fi}
\newbox\grsign \setbox\grsign=\hbox{$>$}
\newdimen\grdimen \grdimen=\ht\grsign
\newbox\laxbox \newbox\gaxbox
\setbox\gaxbox=\hbox{\raise.5ex\hbox{$>$}\llap
     {\lower.5ex\hbox{$\sim$}}}\ht1=\grdimen\dp1=0pt
\setbox\laxbox=\hbox{\raise.5ex\hbox{$<$}\llap
     {\lower.5ex\hbox{$\sim$}}}\ht2=\grdimen\dp2=0pt
\def\gax{$\mathrel{\copy\gaxbox}$}
\def\lax{$\mathrel{\copy\laxbox}$}

\begin{document}

\title[]{Gamma-Ray Bursts \\ \large{Version: \today}}

\author{C. Kouveliotou, S.E. Woosley, L. Piro}

\maketitle

\author[J. Greiner]{Jochen Greiner, 
MPE, Giessenbachstr. 1, 85740 Garching, Germany}

\setcounter{chapter}{5}

\chapter[Multi-wavelength Afterglow Observations]{Discoveries enabled by Multi-wavelength Afterglow Observations of Gamma-Ray Bursts}

\section{Introduction}


The progress in the Gamma-Ray Burst (GRB) field over the last decade and 
prior to the launch of {\it Fermi} mostly occurred in our understanding of 
the afterglow emission and the GRB surroundings. Classical 
observational astronomy, from the radio to X-rays, played a vital role in this
progress as it allowed the identification of GRB counterparts
by drastically improving the position accuracy of the bursters
down to the sub-arcsec level. Once the afterglows were identified, the
full power of optical and near-infrared instrumentation came to play,
and resulted in an overwhelming diversity of observational results
and consequently in the understanding of the properties of the relativistic
outflows, their interaction with the circumsource medium,
as well as the surrounding interstellar medium (ISM) and the host galaxies.
Here we describe the basic multi-wavelength observational properties of 
afterglows, of both long- and short-duration GRBs, as obtained with 
space- (Tab. \ref{sat}) and ground-based instruments. 
The present sample consists of  $\sim$550 X-ray and $\sim$350 optical
afterglows (see http://www.mpe.mpg.de/$\sim$jcg/grbgen.html).


\begin{table}[bh]
\vspace{-0.3cm}
\caption{Main Satellite-Missions contributing to the afterglow sample 
  \label{sat}}
\vspace{-0.1cm}
\begin{tabular}{lcccl}
\hline
Mission/Years  & Instrument & Energy range & Localisation & GRBs\\
\hline
$\!\!$BSAX: 1996--2002 & GRBM      & 2--28 keV & omni-directional &  \\
                       & WFC & 2--28 keV  &  some arcmin &  $\sim$30/yr   \\
$\!\!$HETE-2: 2000--2006$\!\!$  & FREGATE & 6--400 keV & omni-directional &  \\
                         & WXM  & 2--25 keV &  10 arcmin &  $\sim$10/yr   \\
$\!\!$INTEGRAL 2001--    & ACS  & $>$80 keV & omni-directional &  \\
                         & ISGRI& 20--150 keV & 3 arcmin &  $\sim$10/yr   \\
$\!\!$Swift: 2004--      & BAT     & 15--150 keV & 3 arcmin &  $\sim$100/yr  \\
$\!\!$AGILE: 2007--      & $\!\!$SuperAGILE$\!\!$ & 10--40 keV& 5 arcmin & $\sim$6/yr   \\
$\!\!$Fermi: 2008--      & GBM     & 8--30000 keV & some deg & $\sim$250/yr \\
                   & LAT     & 0.1--300 GeV & some arcmin & $\sim$7/yr \\
\hline
\end{tabular}
\vspace{-0.2cm}
\end{table}

\section{Early searches for transient optical emission}

Over the first two decades after the discovery of GRBs (until 1996),
GRB localizations were either  {\it delayed but accurate}, e.g., with
arcmin accuracy, as provided by the Interplanetary Network ({\it IPN} 
\citep{hur95} with typical delays of days or {\it rapid but rough}, e.g., 
within minutes after the GRB trigger, but with at least 2\degs\ error circles
as provided by the BATSE Coordinate Distribution Network  system \citep{bbc96}.

Correspondingly, several alternative strategies were pursued:
(1) searching for quiescent emission in well-localized error boxes
  (assuming the existence of quiescent persistent GRB sources),
(2) {\it post facto} correlating optical monitoring observations temporally 
  overlapping with GRB triggers, and 
(3) quick follow-up observations after a GRB trigger.

\subsection{Searching for persistent quiescent GRB emission}

Archival searches for {\it optical} transients in small GRB error boxes using
large photographic plate collections were initiated at Harvard Observatory 
\citep{sbb84}, and 
then performed at several other observatories 
\citep{hbw87,gfw87}.
Though more than 130 thousand plates were investigated
(see Tab. \ref{plates})
and several optical transients were found, no convincing 
GRB counterpart was identified except the 2008 report on GRB 920925C
\citep{det08}.

The first search for {\it quiescent X-ray} sources in 5 GRB error boxes was
conducted with the {\it Einstein} \citep{piz86}
and {\it EXOSAT} satellites \citep{bo88}.
\cite{gbk95} extended these searches to the 
{\it ROSAT} all-sky-survey data for more
than 30 (15) GRB error boxes determined with
the 2$^{nd}$ (3$^{rd}$)
{\it IPN} catalogs. 
While a number of X-ray sources were found,  
their identification did not reveal any unusual associations,
thus none of these X-ray sources was considered a quiescent GRB
counterpart. 

\begin{table}[hb]
\caption[]{Archival Search for GRB optical counterparts}
\begin{tabular}{lcccc}
\hline
   Group & Observatories & No. of GRB  & No. of & monitoring   \\
         &               & error boxes & plates & time (yrs) \\
\hline
    Schaefer \etal   &  Harvard     & 16 & 32000& 4.25\\
\noalign{\smallskip}
    Hudec \etal      & Ond\v{r}ejov & 21 & 30000&  10\\
\noalign{\smallskip}
    Greiner \etal    & Sonneberg    & 15& 35000 & 2.6\\
\noalign{\smallskip}
    Moskalenko \etal & Odessa       & 40 & 40000 & 1.3  \\
\noalign{\smallskip}
    Schwartz \etal   & S. Barbara & 7 & $\!\!$photoelectric$\!\!$ & 0.1 \\
\hline
\end{tabular}
\label{plates}
\end{table}

\subsection{{\it Post-facto} correlation analysis}

Historically, the hunt for GRB counterparts began with the systematic search 
in photographic exposures serendipitously taken during the
burst event \citep{gwm74}.
This correlation approach was later extended substantially, 
and was also done in a variety of passbands, including
scanning observations with the Cosmic Background Explorer ({\it COBE}) \citep{bon95}.

In the optical band, regular, wide-field sky patrols of two kinds 
were  correlated with GRBs detected with the 
{\it Compton Gamma-Ray Observatory (CGRO)}/ Burst And Transient Source Experiment (BATSE): (i) the Explosive Transient Camera (ETC) exposures
with  a total field of view (FoV) of  40\degs$\times$60\degs\, \citep{vkr95}, 
and (ii) the logistic network of photographic patrols performed at a dozen
observatories worldwide \citep{gwh94}.
During over 4 years
of operation there were five cases when a BATSE GRB occurred during
an ETC observation within or near an ETC FoV.
No optical transients were detected, resulting in
upper limits for the fluence
ratio of gamma to optical luminosities, L$_\gamma$/L$_{opt}$ $\ge$ 2--120.
Unfortunately, in all cases of simultaneous exposures only a part (20\%--80\%)
of the rather large BATSE error box ($>2\degs$; see also Chapter 3) was covered. 
The correlation of BATSE GRBs with photographic wide-field
plates of a network of 11 observatories identified simultaneous plates
for nearly 60 GRBs, with typical limiting magnitudes of m$_{lim}\approx$2--3 mag for an 1\,s duration flash \citep{gwh94}. These limits would correspond to a m$_{lim}\approx$11--12 mag for the canonical afterglow durations discovered in the {\it Swift } era (see also Chapter 5).
Blink comparison of these plates did not reveal any
optical transient (but it did find several new variable stars)
resulting in limits for the flux ratio of gamma-rays to optical emission of 
F$_\gamma$/F$_{opt}$ $\ge$ 1--20.
We know today that these limits were too high for the detection of a canonical optical afterglow.
Instead, the non-detection of an optical counterpart for nearly 60 GRBs is consistent with very bright 
afterglows like e.g.,  GRB\,990123 and 080319B being very rare, of order 1-2\%
of the total afterglow population.

\subsection{Rapid follow-up observations of GRBs}

Early rapid follow-up observations were done already well
before the discovery of afterglows in 1997 (see also Chapter 4), but due to the relatively
large GRB error boxes these searches were not successful in identifying a plausible counterpart.
Already in the early 90s, these rapid follow-up observations
relied on the BAtse COordinate DIstribution
NEtwork (BACODINE), which computed and distributed coordinates of bright GRBs (which had smaller error boxes)
within typically 5 sec after the GRB trigger to interested observers 
\citep{bbc96}.
However, for the few bursts within the FoV of the imaging {\it CGRO}/COMpton TELescope (COMPTEL)
coordinates were determined with much better accuracy and were distributed
after typically 15--30 minutes after detection via the BATSE/COMPTEL/NMSU network \citep{kip94}.


Both, optical and X-ray follow-up observations were performed in the 90's. Notably, GRB\,940301 
was observed seven hours after the GRB trigger with the 1\,m Schmidt telescope at Socorro reaching a limiting magnitude
of m$_{V}$ $\approx$ 16 mag 
\citep{hmp95}; no optical transient was detected. 

{\it ROSAT} pointed observations were initiated within 4 weeks of two GRBs, 
namely GRB\,920501 
\citep{li96}
 and 940301, which due to their close locations had been dubbed
 ``{\it COMPTEL\/} repeater'' GRB\,930704/ 940301 \citep{gbh96}.
None revealed a fading X-ray counterpart.

\section{The {\it BeppoSAX} afterglow discovery}

\begin{figure}[th]
\includegraphics[width=4.8cm, bb=40mm 110mm 180mm 260mm, clip]{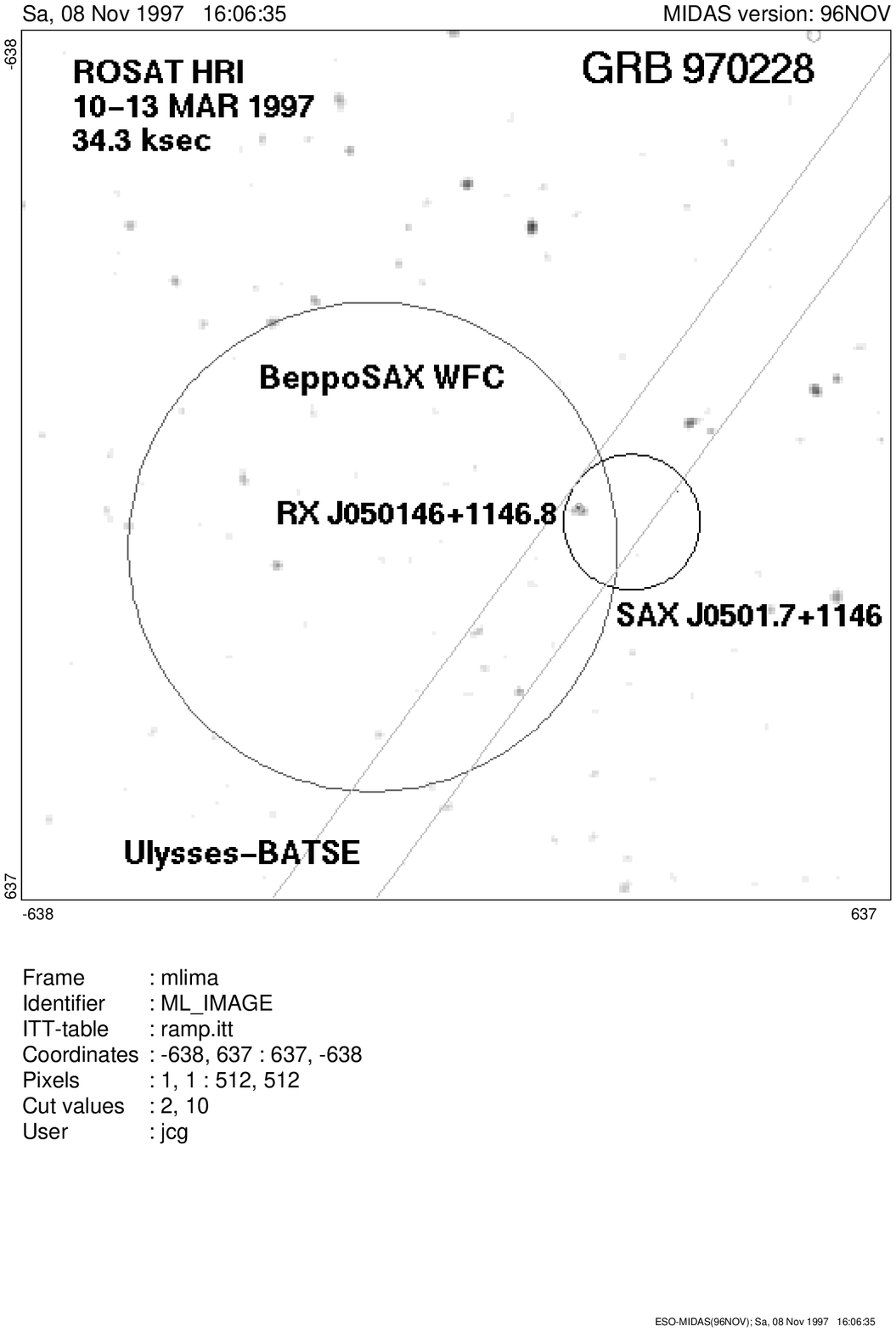}
\includegraphics[width=5.6cm, bb=10mm 118mm 100mm 210mm, clip]{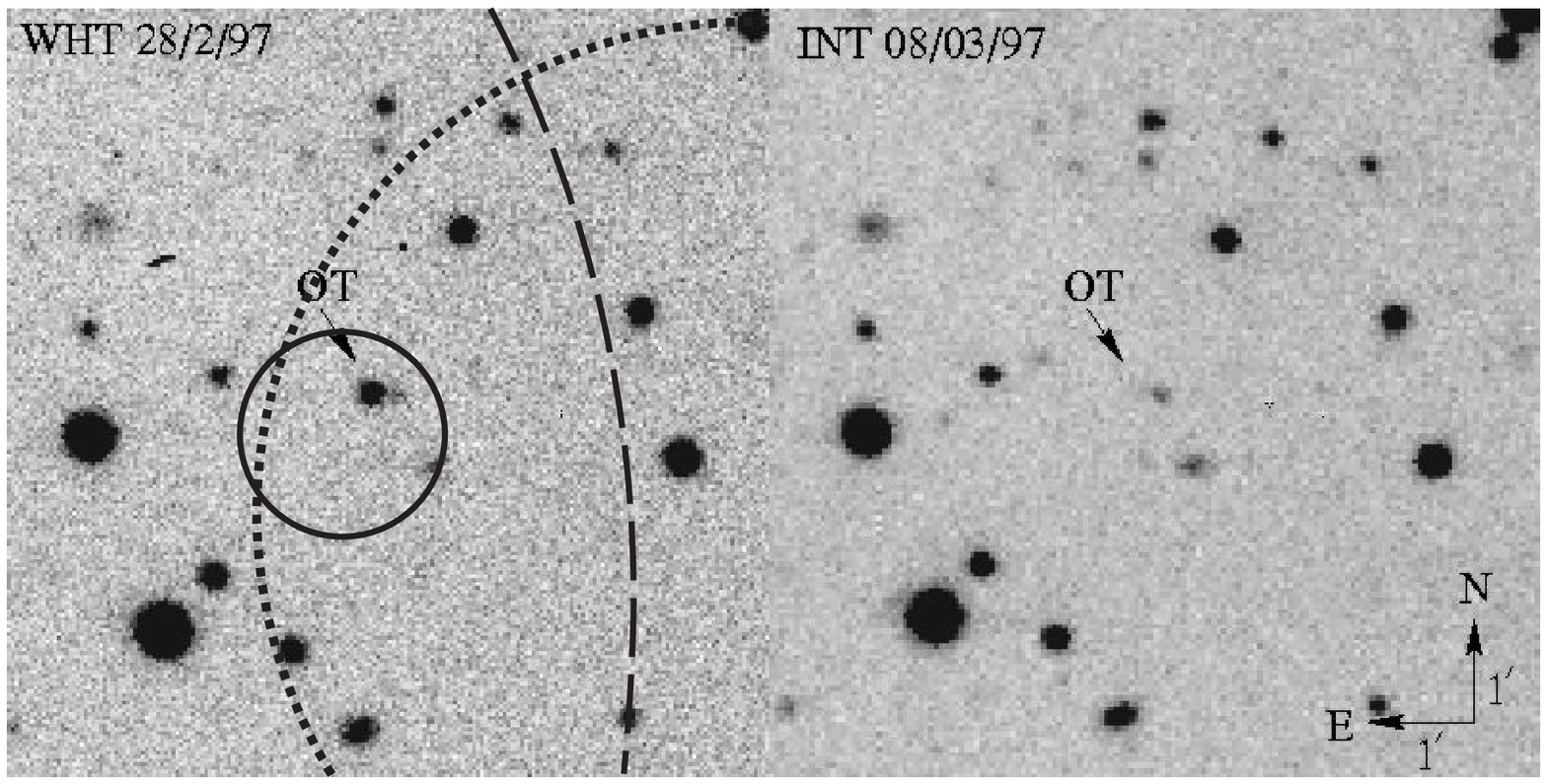}
\caption{Sequence of error circles from $\gamma$-rays to optical
for GRB\,970228, the first GRB for which long-wavelength afterglow
emission was identified. Left: The underlying image is from a 34 ksec
{\it ROSAT}/High-Resolution Imager (HRI) observation \citep{fga98}, with the large circle
showing the 3$\sigma$ error circle of the X-ray afterglow as determined with
the {\it BeppoSAX}/Wide-Field Camera (WFC). The smaller circle is the
$\approx$1~arcmin error circle of the fading source SAX\,J$0501.7+1146$ 
found with the two {\it BeppoSAX}/Narrow-Field instrument (NFI) pointings, and the two
straight lines mark the triangulation circle
derived from the {\it BeppoSAX} and {\it Ulysses} timings  \citep{hur97}.
Right: Optical image taken on 1997 February 28 \citep{vp97}
at the William Herschel Telescope
(Canary Islands) with the WFC error circle
marked as a dashed segment, the NFI error circle with the dotted segment,
and the 10\asec\ {\it ROSAT}/HRI error box as a full circle. The optical
transient (OT) falls right into the {\it ROSAT}/HRI error box. }
\label{grb970228}   
\end{figure}

The launch in 1996 of the Italian-Dutch Satellite per Astronomia X, {\it SAX}, 
ushered a major breakthrough in our understanding of GRBs (for a detailed 
description of the {\it SAX} results see also Chapter 4). Its unprecedented 
localization accuracy ($\sim5$\amin; 2--35 keV), rapid notification 
(within minutes of the GRB) was coupled with its fast slewing capability 
(a few hours) and repointing with its co-aligned narrow field X-ray telescopes.
Despite the fact that only $\approx$3.5\% of its total  observing time
(or 1.5\% of all observations) was spent on GRBs, {\it BeppoSAX} brought a
revolution in the field of GRBs allowing the tools of 
optical/NIR/radio astronomy to be applied to these fascinating objects.

The follow-up observation of GRB\,970228 led to the discovery of the first
X-ray and optical afterglow 
(Fig. \ref{grb970228}) \citep{cos97, vp97}. 
The next important event, the rapid localization of GRB\,970508, allowed the first
measurement of the GRB distance scale via optical spectroscopy.
GRB\,970508 was also the burst with the first radio afterglow. These multi-wavelength observations
provided the first 
observational evidence for the fireball scenario \citep{mdk97, fkn97}.
Subsequent measurements within the next two years demonstrated the
extragalactic nature of GRBs through more redshift measurements of the optical 
afterglow emission as well as of the host galaxies,
and firmly established GRBs as the most luminous objects known in the Universe.
The year 1998 also saw the discovery of GRB\,980425, which was subsequently
associated with a supernova (SN\,1998bw) \citep{gvp98}.

During its lifetime, {\it BeppoSAX} observed 56 GRBs and slewed to 
36 of these \citep{pis04} within typically 5--24 hrs (average around 8 hrs).
X-ray afterglows were discovered  in over 90\% of the cases
and their fundamental properties were established.
It was found that the X-ray flux fades with a power law dependence 
$t^{-\alpha}$, with $\alpha \sim 1.4$ \citep{piro01}.
The X-ray spectrum is well described with a power law $\nu^{-\beta}$of
of slope $\beta \sim 0.9$. The observed absorption is, within the errors, 
always compatible with the Galactic foreground absorption. The observed
flux at a given time after the burst, which is proportional to 
$(1+z)^{\beta - \alpha}$, shows a pretty narrow distribution, since the cosmological spectral redshift (K correction) 
and temporal decay roughly compensate each other: the mean flux in the
$1-10$ keV band at 11 hrs after the burst is about 
$5\times 10^{-13}$ erg cm$^{-2}$ s$^{-1}$ \citep{piro01}.
The overall energy emitted in this late afterglow phase ($>6-8$ hrs)
is typically a few percent of the GRB energy.

We shortly note here two major results because {\it BeppoSAX}
laid the foundations for their studies:
(i) Jet breaks: Being a geometrical effect, jet breaks in afterglow light 
  curves are achromatic, and indeed a number of cases with such breaks
  at $0.5-1$ days after the burst were detected. This provided the
  early observational evidence of beaming in GRBs.
(ii) Confirmation of the basic synchrotron scenario: The broad-band
  spectral energy distribution (SED) was predicted to consist of four segments
  with different powerlaw slopes. The breaks in the SED were found only
  in very few cases, first in GRB\,970508 \citep{wig99}, but provided the first
  observational evidence of a rather low circumburst density (0.03 cm$^{-3}$)
  and a large equivalent isotropic energy (3$\times$10$^{52}$ erg).
Further details on both topics are given in Chapters 8 and 11.

\section{Multiwavelength observations}

The detection of the first optical afterglow(s)
sparked an international observing effort, which was unique,
except perhaps for SN\,1987A. All major ground-based telescopes were
used at optical, infrared as well as in radio wavelengths, and basically every
space-born observatory since then has observed GRBs.  The {\it HETE-2}
satellite \citep{ric02}, launched in October 2000, continued to provide
rapid and arcmin sized GRB localizations at a rate of about 2 per month
after {\it BeppoSAX} had been switched off in April 2003. {\it Swift}, launched
in November 2004, revolutionized our knowledge on the afterglow phenomena.
Over the last 13 years (February 1997 -- June 2010) a total of 870 GRBs
have been localized within a day to less than one square-degree size error boxes,
and X-ray afterglows have been detected basically for each of those bursts
for which X-ray observations have been done within a few days
(see http://www.mpe.mpg.de/$\sim$jcg/grbgen.html).

 \subsection{Contemporaneous, prompt multiwavelength emission}

\begin{table}[ht]
   \caption{Top 10 brightest optical afterglows of GRBs.  
    Another 19 afterglows reached a maximum brighter
   than 15 mag in one color.} 
      \begin{tabular}{lrccl}
      \hline
GRB     &  $\!\!$Brightness  & Filter  & Time after    &     Reference \\
        &       (mag)~~   &            &  GRB (sec)   &               \\
      \hline
080319B &     3.8~~   &  K$_s$   &  65   &  \cite{bpl09}  \\
080319B &     5.4~~   &  V       &  53   &  \cite{rks08}  \\
990123  &     8.9~~   &  white   &  50   &  \cite{abb99}  \\
061007  &     9.9~~   &  white   &  94   &  \cite{raa09}  \\
060117  &    10.1~~   &  R$_c$   &  129  &  \cite{jpk06}  \\
060418  &    10.2~~   &  K$^\prime$& 168  &  \cite{mvm07}  \\
061126  &    11.0~~   &  K$_s$   &  137  &  \cite{pbb08}   \\
081203A &    11.6~~   &  I$_c$   &  415  &  \cite{wmb08}   \\
081121  &    11.6~~   &  white   &  60   &  \cite{yur08}$\!\!$   \\
090102  &    11.8~~   &  H       &  102  &  \cite{gkp10}   \\
030329  &    11.9~~   &  J       &  8100 &  \cite{nhk03}  \\
      \hline
   \end{tabular}
   \label{brightAG}
\end{table}

Some optical afterglows have shown substantial variability
at early times. One can distinguish
a component which tracks the prompt gamma-rays
(GRB\,041219A \citep{vest05, bbs05}, GRB\,050820A \citep{vest06},
 GRB\,080319B \citep{rks08})
and an afterglow component which starts during or shortly after the
prompt phase
(GRB\,990123 \citep{abb99}, 021211 \citep{lfc03}, 
GRB 060111B \citep{kgs06}).
The former component has been attributed to internal shocks,
while the latter was interpreted as reverse shock emission,
e.g. \cite{sap99, mer99}.
The internal shock emission is relativistic,
and the timescales in the observer frame are shortened by
$\Gamma^{-2}$, with $\Gamma$ being the bulk Lorentz factor
which typically is assumed to be of order 300--500.
The reverse shock is predicted to happen with little delay with respect to
the gamma-ray emission (unless the Lorentz factor is very small), and
the corresponding optical emission decays with a power law index of $2$
for a constant density environment, or up to $2.8$ for
a wind density profile \citep{kob00}.

 \subsection{Dark bursts \label{darksec}}

Originally, those GRBs with X-ray afterglows but without optical detection
(about 50\%) were coined as ``dark GRBs''. The ``darkness'' in the optical 
was assumed to be due to one (or more) of several reasons \citep{fyn01}: 
the afterglow could
(i) have an intrinsically low luminosity, e.g., due
  to a low-density environment or low explosion energy,
(ii) be strongly absorbed by intervening material, either very
  local around the GRB, or along the line-of-sight through the host galaxy,
or (iii) be at high redshift ($z>6$) so that Ly$\alpha$
  blanketing and absorption by intervening Lyman-limit systems  would prohibit 
 detection in the $R$ band (most frequently used in the optical).
An analysis of a subsample of GRBs, namely those with particularly
accurate positions provided with the Soft X-ray Camera on {\it HETE-2},
showed that optical afterglows were found for 10 out of 11 GRBs
\citep{villa04}. This suggested that the majority of dark GRBs are
neither at high redshift
nor strongly absorbed, but just faint, i.e., the spread in afterglow
brightness at a given time after the GRB is much larger than
previous observations had indicated.
However, since 2004 the {\it Swift} observations have provided a plethora of 
locations at the few arcsec level within minutes of the GRB, and the fraction
of dark bursts is still above $\sim$30\%. 


Very recently, a sample with nearly complete afterglow detections was reported,
which had been created by selecting those GRBs for which observations with the
Gamma-Ray Burst Optical/Near-Infrared Detector {\it GROND} (operated at the 
2.2m telescope at the La Silla Observatory
\citep{gbc08}) started within 30 min after the burst \citep{gkk10}.
With a 95\% detection completeness and a simultaneously obtained 7-band
spectral energy distribution for all these bursts, rest-frame extinction 
$A_{\rm V}$ is accurately measured for the first time in a coherent way. 
Substantially more bursts with $A_{\rm V} >0.5$ mag are found than in 
previous samples \cite{kkz10}, and in 
many cases a moderate redshift (in the 1--3 range) enhances the effect
in the observer frame. The properties of this sample demonstrate that
the darkness can be explained by a combination of (i) moderate extinction
at moderate redshift, and (ii) a  ($\sim$10\%) fraction of bursts
at redshift $z>5$. This strengthens similar earlier suggestions 
\citep[e.g.][]{ckh09, pcb09},
which were based on a combination of early detections and host galaxy
studies of the non-detected afterglows.

 \subsection{Spectral lines}

Line detections have been reported at optical and X-ray wavelengths.
These early X-ray line detections were based on {\it BeppoSAX}, {\it ASCA},
{\it Chandra} and XMM-{\it Newton} observations. A comprehensive analysis, 
however, of $>$200 {\it Swift} bursts did not reveal any significant 
X-ray lines \citep{rcm08, hvo08}.
Therefore, in the following we will constrain ourselves to optical lines.

\begin{figure}[th]
\includegraphics[width=4.5cm]{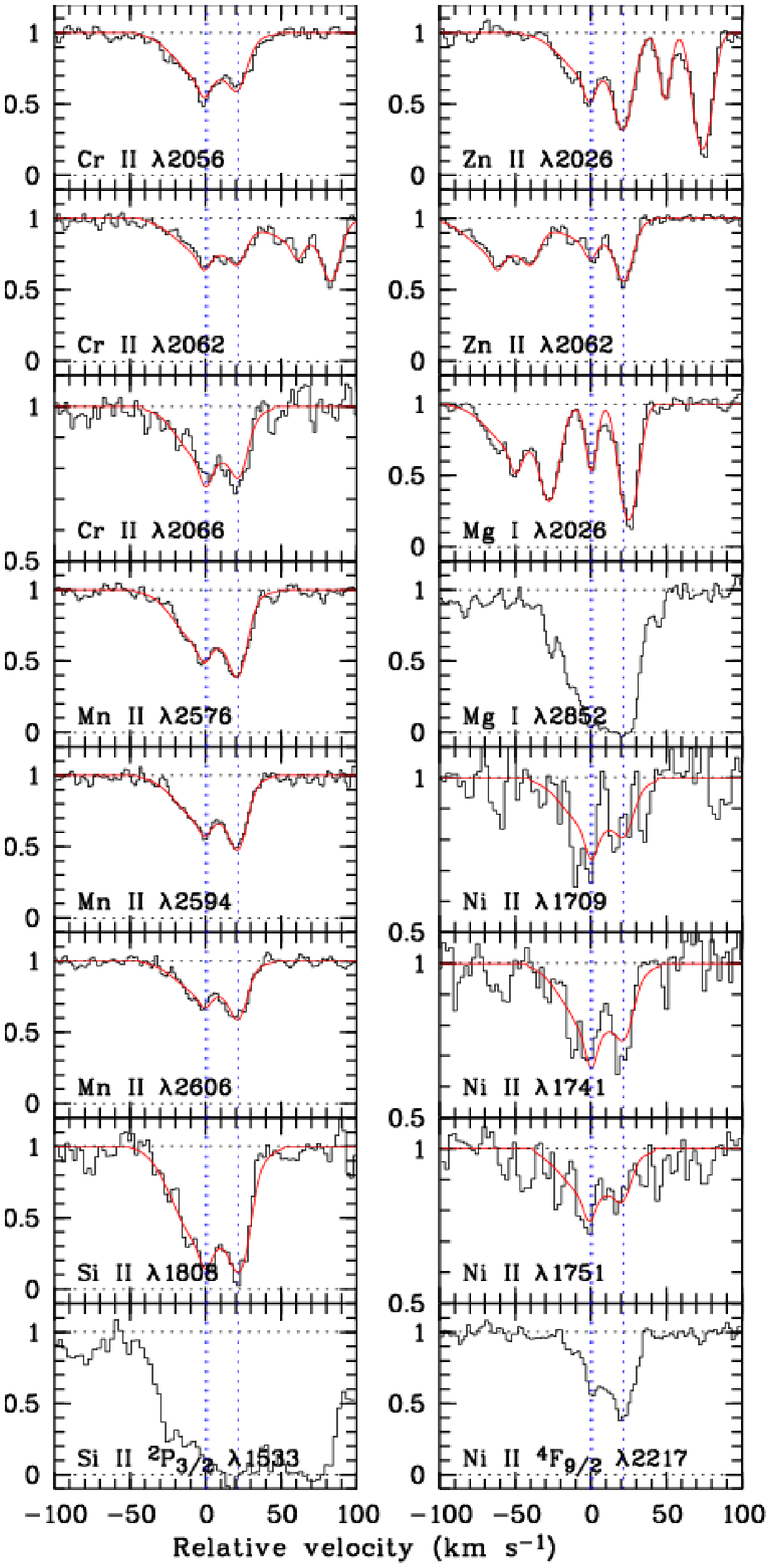}
\includegraphics[width=6.4cm]{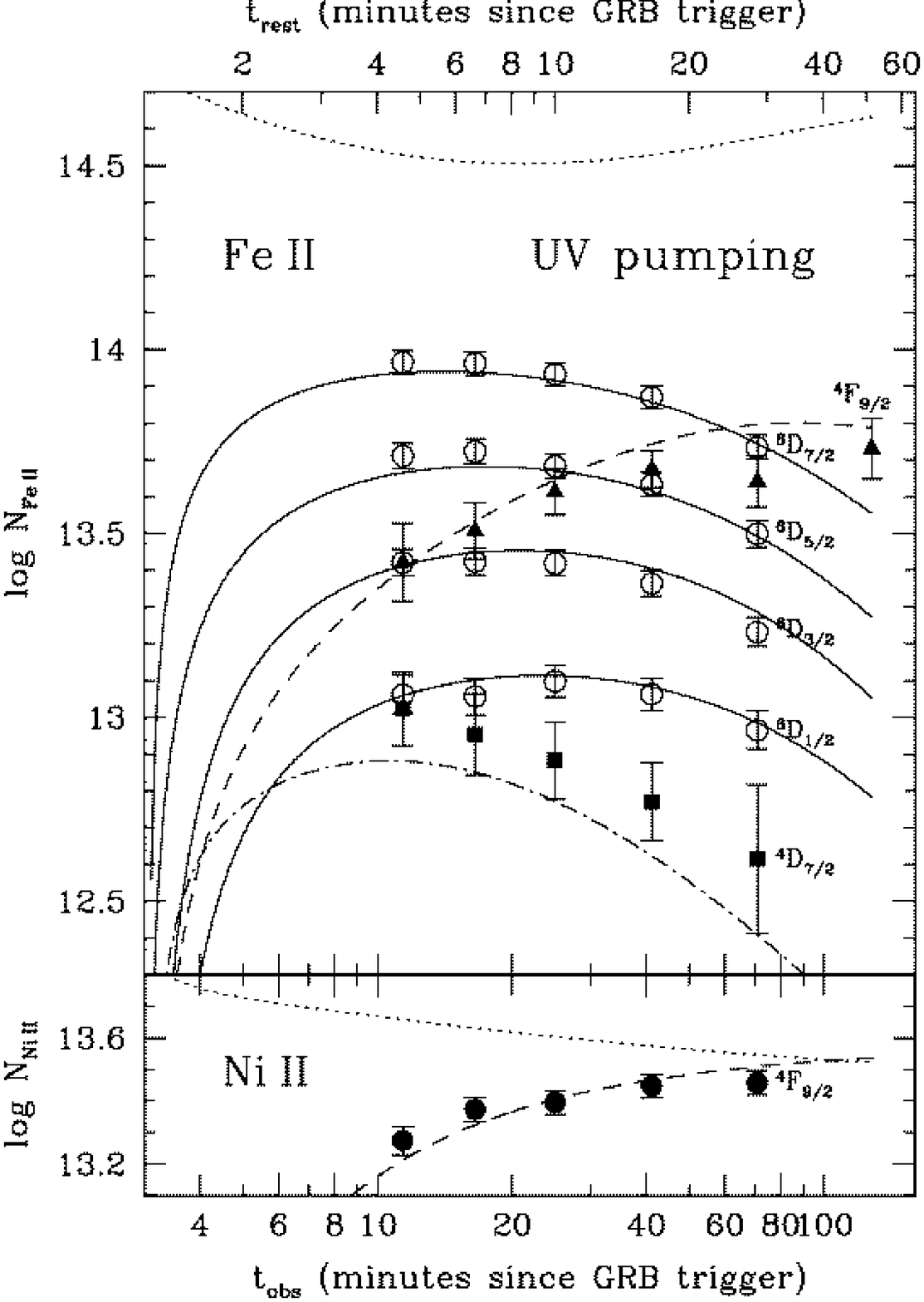}
\caption{{\bf Left:} Absorption line profiles for a variety of transitions 
    detected at the GRB\,060418 redshift. Red lines show the Voigt-profiles
    fits of low-ionization species. 
   {\bf Right:} Observed total column densities for the fine-structure lines 
  (open circles), the first metastable level (filled triangle) and the
   second metastable level (filled squares) of FeII (top) and the total 
   column densities for NiII (bottom). Lines are the best-fit UV pumping
   model.
From \citet{vls07}}.
\label{UVpump}   
\end{figure}

Optical/NIR spectroscopy of afterglows usually reveals absorption lines
of (typically more than one) system along the line of sight between the 
GRB and the observer. The system with the largest redshift is then assigned 
to be the redshift of the GRB. Formally, these absorption redshifts
are still a lower limit, but one would have to assume a contrived empty 
environment if the GRB were at a much larger redshift than the last absorption
system and if it would leave no measurable imprint in the spectrum.
Moreover, the detection of a Lyman cutoff or (even time-variable) lines from
fine-structure levels provide stringent limits.
Measurements of the equivalent widths of the absorption features 
(e.g., Fig. \ref{UVpump}) allows us to derive column densities of metal lines
and neutral hydrogen, 
as well as the metallicity and dust content along the line of sight. 
A special case are absorption lines from fine-structure and other metastable
levels of ions such as O$^o$, Si$^+$ and Fe$^+$, which are ubiquitous in 
GRB-Damped Lyman $\alpha$ systems (DLAs) \citep{vel04, cpb05, bpc06, pcb06}, 
and very likely are excited by the GRB emission.

Interestingly, in about half of the cases, the GRB-DLAs exhibit column 
densities of log($N_H$) $\sim10^{22} cm^{-2}$ or above \citep{fjp09}. 
This is in contrast to
the only few such systems in QSO-DLAs \citep{nlp08}, and is likely due
to the smaller range of galactocentric distances probed by GRB sightlines 
(for a detailed discussion on GRB-DLAs see also Chapter 13). 

An interesting puzzle was brought up by \citet{pro06}, namely that the
number density of strong (equivalent widths $>$ 1 \AA) 
intervening Mg{\sc ii} absorbers detected in 
GRB afterglow spectra at redshifts $0.5<z<2$, is nearly 4 times larger than 
those in the QSO spectra.
Similar analyses based on a different dataset found a factor 2 larger
incidence rate (with higher significance than the earlier factor 4), 
but only for strong absorbers, while for weaker absorbers,
(equivalent widths in the 0.3--1.0 \AA\ range), the incidence rate
was consistent with that in QSO spectra 
\citep{vpl09, tlp09}.
A similar study with C{\sc iv} absorbers did not reveal any differences between GRB and QSO
sightlines \citep{ssv07}.
A number of possible explanations have been proposed \citep{pvl07}, including
a dust extinction bias, or different beam sizes of the sources,
or lensing amplification, but none provided a conclusive solution
of this discrepancy so far.

Depending on its brightness, early host spectra might already 
show emission lines. Usually, host spectroscopy is done when the
GRB afterglow has faded away. To date (2010), there has not been a single 
case of emission lines being at a larger or smaller redshift than the 
highest-redshift absorption system, supporting the assumption that the GRB 
belongs to the corresponding host galaxy.
Besides giving the redshift, the observed emission lines of O{\sc ii}, 
[O{\sc iii}] and
the Balmer series are used to infer the global extinction, metallicity
and star formation rate \citep{sgb09, lbk10}.
Note that in this case these are host-integrated quantities, in contrast
to the line-of-sight measurements via absorption lines.

   \subsection{Line Variability}

Variability of both absorption and emission lines has been searched
for, though on substantially different timescales.

Variable absorption lines, involving the fine structure of the ground level
and other metastable energy levels of Fe$^+$ and Ni$^+$ were first
modelled (Fig. \ref{UVpump}) for GRBs\,020813 and 060418 \citep{dzcp06, vls07}. 
It was demonstrated that these lines are formed by UV pumping,
i.e. excitation to an upper level due to the absorption of a UV photon,
followed by de-excitation cascades, as suggested by \citet{pcb06}. 
This interpretation allowed the first determination of the distance 
between the GRB and its DLA to an astonishing 1.7 kpc.
Later re-modelling with a different set of atomic abundances increased
this distance to 2.0$\pm$0.3 kpc \citep{lvs09}. For GRB\,050730, the 
same authors derived a distance (near-side of the cloud) of 440$\pm$30 pc
for a cloud 520$^{+240}_{-190}$ pc size (along the line of sight).
This is in contrast to a distance of only about 50--100 pc for which the
GRB radiation can ionize hydrogen. 
The global picture derived from modelling these variable lines is that
of the absorber being a large, diffuse cloud with a broadening parameter and 
physical size typical of the Galactic ISM, with low metallicity and low
dust content, and at a distance at least 0.1--1 kpc away from the GRB
\citep{lvs09}.

Variability of the emission lines was expected since the GRB prompt and 
afterglow emission ionizes its surrounding to substantial distances. 
Depending on the density of the circumburst medium,
this leads to recombination lines over timescales of years which could
compete with the emission lines usually assigned to star-formation.
In fact, it had been proposed to use the GRB-induced lines 
for the identification of
remnants of GRBs in nearby galaxies \citep{bah92, prl00}.
Such a search was indeed conducted for the host galaxy of GRB\,990712, but no
variability was found in the OIII [5007] line over a timescale of 
6 years \citep{aky06}.

   \subsection{Continuum variability}

   \subsubsection{Early lightcurve behaviour}

The early time GRB afterglow behaviour depends strongly  on the wavelength range
considered. At soft X-rays, {\it Swift} has found surprisingly rapid 
variability 
in both, short- and long-duration GRBs. Yet, many of the early light curves
show a canonical behaviour with three distinct power law
segments \citep{ncg06}: a bright, rapidly declining ($t^{-\alpha}$, with 
$\alpha > 3$) emission, which smoothly connects to the prompt emission
both temporally and spectrally \citep{tgc05, bcg05}, 
followed by a steep-to-shallow 
transition, which is usually accompanied by a change in the power-law index
of the spectrum. The first break has been interpreted \citep{ncg06, zfd06}
as the slowly decaying forward shock emission as it becomes dominant over the 
rapidly declining tail emission of the prompt
$\gamma$-rays as seen from large angles \citep{kup00}.
The subsequent shallow phase is commonly interpreted as due to continuous
energy injection into the external shock \citep{ncg06, zfd06}, which implies
that most of the energy in the afterglow shock was either injected at 
late times after the prompt $\gamma$-ray emission, or was originally in 
slow material that would not have contributed to the prompt emission.
This shallow phase then transitions into the late afterglow phase
with no clear evidence for a spectral change.

\begin{figure}[th]
\includegraphics[width=\textwidth]{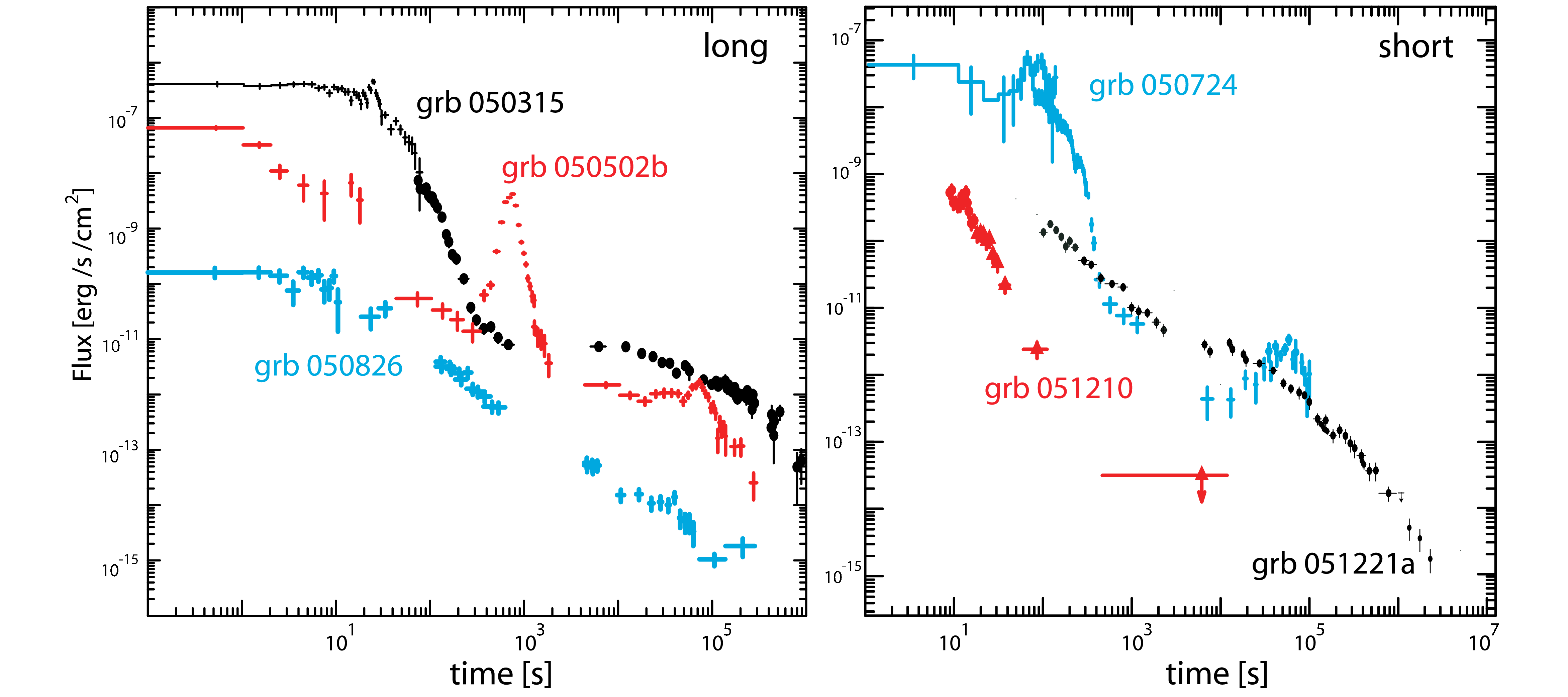}
\caption{Representative examples of X-ray afterglow light curves of long (left)
  and short-duration (right) GRBs.
 From \citet{grf09}.}
\label{XAG_lc}   
\end{figure}

\begin{figure}[th]
\includegraphics[angle=270,width=\textwidth]{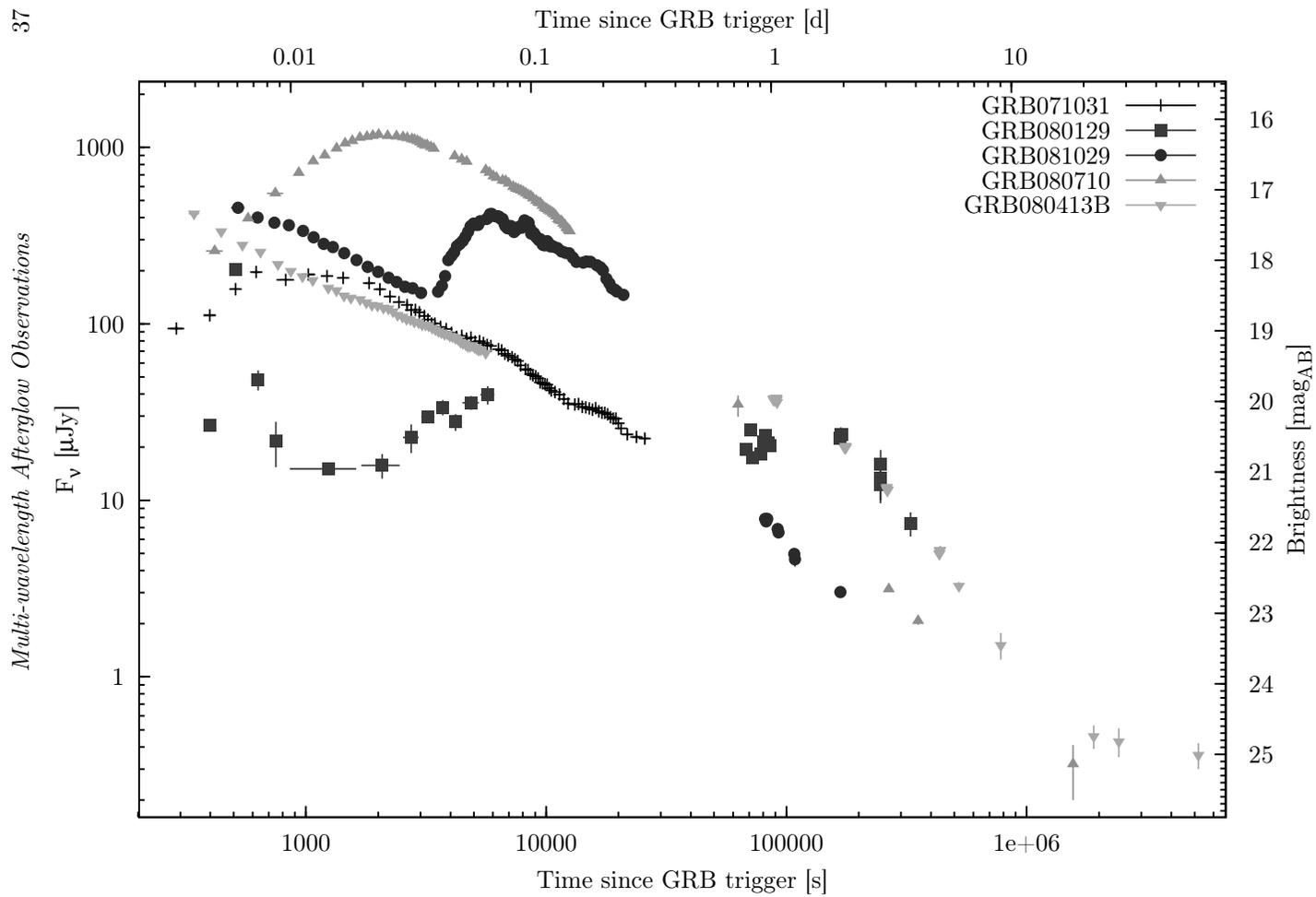}
\caption{Representative examples of optical light curves of long-duration GRBs
  as measured with GROND. The light curve diversity is similar to that in 
  X-rays.
 }
\label{OAG_lc}   
\end{figure}

Extended emission lasting about 100 sec has been detected at hard X- and 
gamma-rays  in about 25\% of the short bursts \citep{nob06}. 
Though these tails were known already from {\it HETE-2} \citep{vlr05} and 
{\it CGRO}/BATSE \citep{lrg01, conn02, nob06}
a systematic study was only possible with {\it Swift} \citep{ngs10}, since this
emission is rather soft and has spurred debate on whether it is afterglow 
or prompt emission.

The optical afterglow behaviour is at least as diverse as the X-ray one 
(Fig. \ref{OAG_lc}):
A large fraction of the afterglows show the canonical smooth power law decay,
but some show a completely different behaviour. There are rare cases
(like GRBs 990123 or 080319B) which are dominated by very bright, fast 
decaying emission, which is usually interpreted as the appearance of the 
reverse shock
(\cite{abb99, rks08, bpl09}, but see e.g., \cite{geg09} for an alternative 
interpretation).
About 10--20\% of the optical afterglows exhibit an increase in their
brightness during the first few hundred seconds. This has been observed both 
with the {\it Swift}/ Ultra Violet Optical Telescope (UVOT) \citep{ops09} 
as well as with fast-slewing telescopes from the 
ground \citep{rsp04, qry06, yar06, mvm07, kkg08, cdk08, kgm09}. No color 
evolution, however, was seen during the rise and the turn-over towards decay.
The deceleration of the forward shock by the external
medium has been favoured as an explanation for light curve shapes, where the 
rise and subsequent decay can be modelled with a broken power-law. The time 
of maximum light
was also used to derive initial Lorentz factors of 80--300.
Cases, where the decay showed another break at early times, and the
power-law indices of the rise and first decay did not match the standard
fireball prediction, have been interpreted as a signature of jet emission 
seen off-axis \citep{panaitescu08}.
Assuming that the jet structure has a power-law angular distribution,
there is a correlation between initial rise time and the slope of the
first fading after maximum, which can explain the observed diversity of
light curves \citep{panaitescu08} -- though applications to larger
samples have not confirmed this trend \citep{kba09, kkz10}.

An interesting consistency check is now possible with measurements from
{\it Fermi} and {\it INTEGRAL}: the Lorentz factor can also be determined 
from the
variability of the gamma-ray emission \citep{lis01}. For GRB\,080928
\citep{rosk10} this comparison has been attempted for the first time, and the 
two values of the initial Lorentz factor are indeed broadly consistent.

     \subsubsection{Jet-breaks}

An observer will detect emission due to relativistic beaming of the emission 
from the GRB blast wave within an angle $\sim$1/$\Gamma$ of the line of sight 
(see also Chapter 11). The afterglow is thus a signature of the geometry of 
the ejecta. Until the blast wave has decelerated such that its opening
angle is $\sim$1/$\Gamma$, its gradual fading
is partly compensated by an increasing emission region. Only at angles larger 
than 1/$\Gamma$,
does the observed emission decay with a power-law index of $>$2 and can be 
described under a spherically symmetric model.
Since this transition is a geometric effect, the slope change in the 
afterglow light decay should be achromatic \citep{rho99}, that is observable 
at all wavelengths at the same time.

In the pre-{\it Swift} era, this achromatic steepening was commonly
reported in the optical afterglows and interpreted as the
indication of beamed emission. Using the pre-{\it Swift} data, collimation 
factors of 
$\Omega$/4$\pi$ \lax 0.01, corresponding to half opening angles of \lax 8\degs\,
were derived from the timing of these breaks \citep{fks01, bfk03}.

However, with {\it Swift} only a small fraction of bursts has
been reported with convincing evidence for an X-ray jet break \citep{rlb09}.
Today a general consensus has developed according to which the breaks in 
{\it Swift}-detected bursts occur at later times due to their larger mean redshift, 
and thus at flux levels  beyond the sensitivity of standard follow-up campaigns. Recent results
of a dedicated long-term monitoring of X-ray afterglows with {\it Chandra}
seems to recover jet breaks for about 40\% of the {\it Chandra} observed bursts
\citep{bur10}.

     \subsubsection{X-ray flares}

GRB\,050502B provided one of the first examples of the dramatic X-ray flaring
activity in the early afterglow evolution \citep{brf05, fbl06}. This burst 
also demonstrated that X-ray flares (measured up to 10 keV) can contain 
energy comparable to the one emitted during the prompt GRB phase in the 
15--300 keV band. Surprisingly, X-ray flares
have been seen in long- and short-duration GRBs, as well as at low and
high redshifts: even GRB\,090423 at z$\sim$8.2 exhibited a flare with
rather standard properties \citep{cmm10}. The majority of the flares
occurs during the first 10$^3$ s after the GRB trigger, but some have also 
been seen as late as 10$^5$ or even 10$^6$ s \citep{cso08} (see next section).
The flares are relatively sharp, with $\Delta t / t \sim 0.1$, and are
spectrally different (harder) than the underlying afterglow emission.
There is considerable spectral evolution during a flare with a
hardening during the rise followed by softening during the decay 
\citep{gpg07, kgm07, gpo07}.
The first case where these flares were seen simultaneous in the
optical/NIR was GRB\,071031 \citep{kgm09}, which showed that the peak 
of the emission shifts at late times from the few keV band into the UV.
Given that the flare phenomenology is very analogous to that of the prompt 
gamma-ray emission, it is now generally accepted that X-ray flares
and gamma-ray pulses are produced by the same mechanism.

     \subsubsection{Early time afterglow features (``humps'') }

Some GRB afterglows (GRBs\,021004, 030329) exhibited ``humps'' on top of the 
canonical optical fading at timescales of 10$^4$-10$^5$ s after the GRB 
onset \citep{lrc02, log04}.
Originally, these humps were interpreted as
the interaction of the blast wave with moderate density enhancements in the
ambient medium, with  a density contrast of order 10 \citep{lrc02}; 
later models employed additional energy injection episodes \citep{bgj04}. 
Optical afterglow variability due to the interaction with the ISM
is not expected later than 10$^6$ s because the
blast wave, once it has swept up enough interstellar material to produce the
canonical afterglow emission, is thought to be only mildly relativistic.
It is possible, but not easy to prove due to lacking X-ray observations, 
that these humps are related to the X-ray flares discussed in the previous
section.

     \subsubsection{Late afterglow features: Supernovae and something else?}

There is now general consensus that the long/soft \citep{kou93} GRBs are 
intimately connected
to the deaths of massive stars. About 70\% of core-collapse supernovae (SNe) 
are those of type II; one of the peculiar sub-classes that form part of the 
other 30\% are type Ib/c supernovae.

\begin{figure}
\includegraphics[width=8cm]{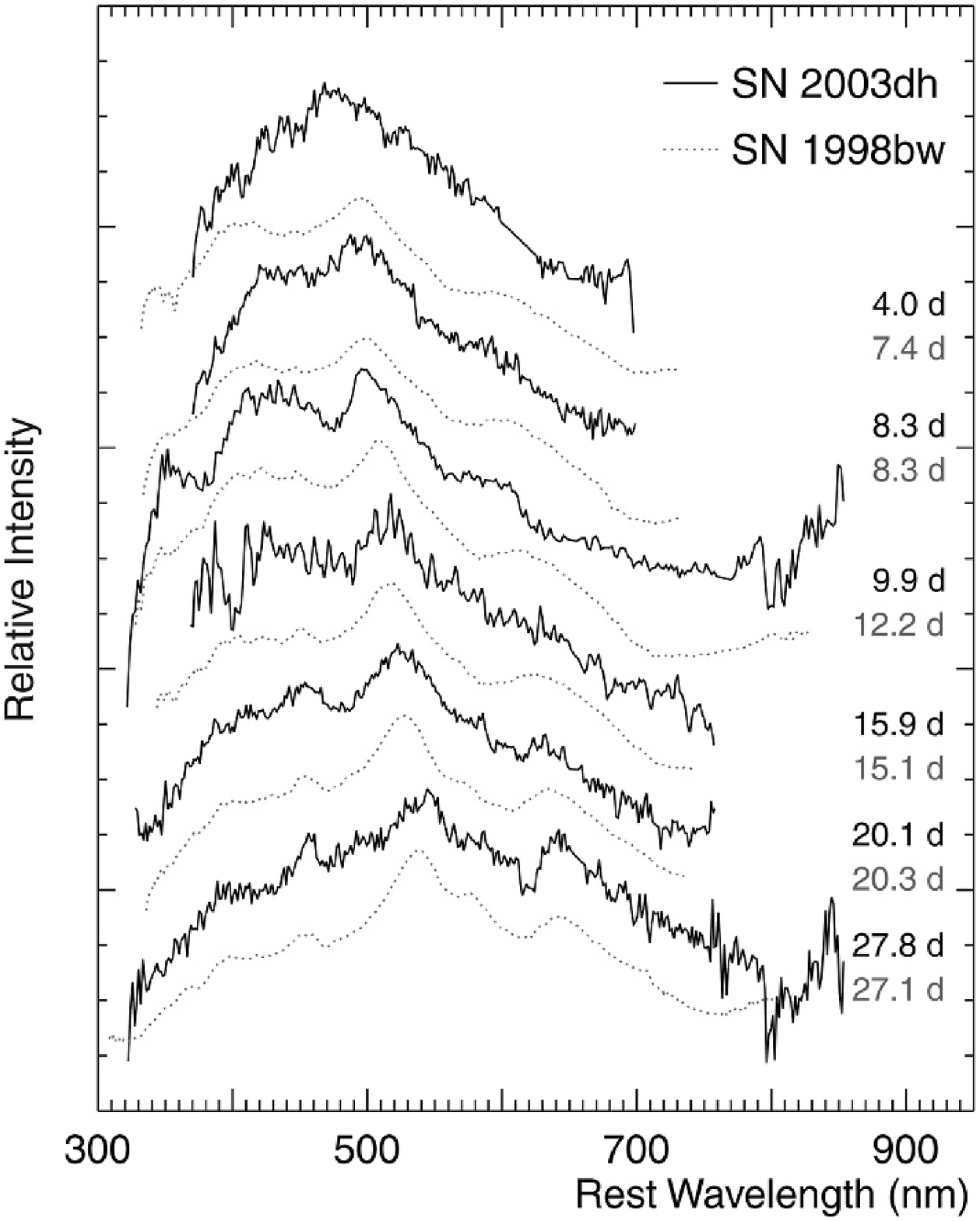}
\hfill\parbox[t]{3.cm}{\vspace*{-6.0cm}\caption[sn]{Evolution
of the optical spectrum of the afterglow of GRB\,030329 showing
features of a SN Ic, as compared to the spectra
of SN\,1998bw (dotted lines). The time after the SN explosion is given on
the right side. From \citet{hjo03} \label{snspec}}}
\end{figure}

While the supernova-GRB connection was proposed some years
ago \citep{gvp98, iwa98}, the unambiguous spectroscopic identification
of the  lowest-redshift long-duration GRBs as supernovae
during the last decade provided convincing
evidence for this association \citep{hjo03,sta03} (Fig. \ref{snspec}).
The supernovae in the five spectroscopically confirmed gamma-ray
bursts (GRB\,980425/SN\,1998bw, 030329/2003dh, 031203/2003lw,
060218/2006aj and 100316D/2010bh) are all of type Ic,
with unusually large kinetic energy (very large expansion
velocities of order 10--30 thousand km/s were measured after 10 days) and
ejected mass of radioactive
$^{56}$Ni; such SNe were called hypernovae by \citealt{pac98}.
Their latter property in particular suggests progenitors
with masses \gax 40 \msun\ \citep{nom04}, though the detailed analysis of
the light curve and spectra of GRB 060218 / SN 2006aj showed that the
initial mass was only $\sim$20 \msun, indicating a possibly broader range
of progenitor masses leading to a GRB \citep{mdn06}.
Theoretically, SNe Ib/c are favoured over type II because
the former have typically smaller envelope masses,
and are thus thought to allow easier break-out of the GRB jet.
Moreover, the lack of hydrogen lines in the GRB afterglow spectra is consistent
with the collapsar model, where the progenitor star lost its hydrogen envelope 
to become a Wolf-Rayet star before collapsing.

In contrast to these relatively similar spectroscopic properties among
the GRB-SN, the $\gamma$-ray emission properties of the corresponding GRBs 
differ in their total emitted energy \citep{krg07}, temporal profile and 
spectral shape, implying that the $\gamma$-ray properties are not determined 
by the progenitor mass, but most likely by completely different properties
\citep{gfp06}.

Two noteworthy exceptions to this picture of SNe detections associated with the
nearest bursts are GRB\,060505 and 060614.  A host galaxy at z=0.125 was 
associated with this burst based on two optical emission lines; moreover 
this was clearly a long-duration burst (T$_{90}$=102 s).
However, no SN was found in the error box of GRB\,060614  \citep{fwt06, dcp06} 
to limits about a factor 100 fainter than previous detections. Both bursts 
have spurred extensive discussions on the homogeneity of
the class of GRB-SN, and the classification of GRBs \citep{gfp06, gnb06, zzv09}.
Gehrels et al. (2006) proposed a third parameter for GRB classification based 
on their spectral lags (difference of arrival times between high and low-energy 
photons) and their peak luminosities. According to this criterion, the spectral 
lag of GRB\,060614 would place this burst entirely within the short-duration 
GRB subclass \citep{gnb06}.

Finally, a number of GRBs show optical humps at late times, but earlier than 
the expected appearances of their related SNe, around $10^5-10^6$ s 
\citep{mkr10}.
Multi-color light curves show that these humps are achromatic, excluding
very early SNe. The cause of these humps is still a matter of debate.

     \subsubsection{Very late afterglow evolution \label{dipank}}

The expansion of GRB afterglows, while being initially ultra-relativistic,
slows down in the course of time and eventually enters a sub-relativistic
phase after several tens to hundreds of days.  
With the notable exception of GRB\,060729 \citep{gbw10},
most afterglows are too faint to be detectable in most wavelengths at such
late times, and their observations are confined mainly to 
low-frequency radio bands.  Despite a large number of afterglow detections
at radio wavelengths \citep[see, e.g.][]{frail+03},  only two well 
studied examples exist so far of observations, in multiple radio bands, 
deep into the non-relativistic phase: 
GRB\,970508 
and GRB\,030329.
For the
former, radio follow-up at the 1.4~GHz to 8.4~GHz range was conducted for
more than 400 days post-burst, while the transition to non-relativistic 
expansion occurred at $\sim 100$~days 
\citep{fwk00}.
In the case of GRB~030329, radio observations at several frequencies, 
(610~MHz to 4.8~GHz) over $\sim 1200$~days after the
burst have been reported 
\citep{horst07}
The non-relativistic
transition time in this case was estimated to be $\sim 60-80$~days.

Observations well within the non-relativistic phase provide a useful additional
tool to derive the physical parameters of the burst, in particular the total (bolometric) 
energy \citep{onp05,krg07}.  
The dynamics in this regime is governed by the Sedov-Taylor 
solution, which is different from the Blandford-McKee solution in the early
relativistic phase before the jet break.  Burst parameters derived from the
non-relativistic phase alone may, therefore, be considered as a set of independent
measurements, which serve as a useful check on the quantities derived 
from the relativistic phase evolution.  Multiband modelling of the relativistic
phase needs to include a description of the angular distribution of the energy and
Lorentz factor of the outflow,
which remains uncertain even in the presence of a well-determined 
jet break.  In the deep non-relativistic phase, however, the expansion
of the blast wave is expected to have become nearly isotropic, so the energy
estimates are much less prone to uncertainties arising from collimation
effects.  The total energy $E_{\rm ST}$ estimated for the Sedov-Taylor 
non-relativistic phase, together with the isotropic equivalent 
energy $E_{\rm iso}$ estimated from 
burst fluence and relativistic phase modelling, provide a useful indicator 
of the degree
of initial collimation of the relativistic outflow.  In the case of GRB~970508
the estimated values of 
$E_{\rm ST}$ and $E_{\rm iso}$ are $\sim 5\times 10^{50}$~erg
and $\sim 10^{52}$~erg, respectively, suggesting an initial collimation angle
$\leq 20^{\circ}$ \citep{fwk00}.  For GRB~030329, the 
corresponding
estimates are $\sim 8\times 10^{50}$~erg and $\sim 7-8\times 10^{51}$~erg,
respectively
\citep{bkf04, frail+05, horst07}.

Several microphysical quantities  may in fact be a function of the dynamical
regime, and hence may not have the same value in the relativistic and the
non-relativistic phase.  These may include parameters such as 
$\epsilon_{\rm e}$, 
the fraction of the total energy resident in relativistic electrons, 
$\epsilon_B$, the fraction of the total energy resident in 
post-shock magnetic field, and $p$, the power-law index of the electron 
energy distribution.  By modelling the relativistic and the non-relativistic
phase evolution separately, one may in principle be able to conclude 
whether these microphysical parameters are indeed different in the two
phases.  Obtaining a complete solution for physical parameters in the 
non-relativistic phase requires the measurement of all three spectral 
breaks, $\nu_{\rm a}$, $\nu_{\rm m}$ and $\nu_{\rm c}$.  Multi-band radio
light curves can be used to determine the first two of these breaks, but
a direct measurement of the cooling frequency in the non-relativistic
phase has not yet been possible, in the absence of high frequency
observations.  As an approximate estimate, one uses the value of $\nu_{\rm c}$
extrapolated from an earlier, relativistic phase to infer the physical
parameters.

Because of this partial lack of information and also the uncertainties 
inherent in the measurement of spectral parameters, it is not yet possible
to state with confidence whether the microphysical parameters are 
indeed different between the relativistic and the non-relativistic phase.
Nevertheless, in both GRB~970508 and in GRB~030329 one finds that
in the non-relativistic phase the energy in relativistic electrons and that 
in the magnetic field are nearly in equipartition 
\citep{fwk00, horst07},
while in the relativistic phase the derived
estimates of $\epsilon_{\rm B}$ tend to be significantly smaller than
those of $\epsilon_{\rm e}$ \cite[see, e.g.][]{pk01, pk02}.

Another important measurement that is made possible by the long-lasting
radio follow-up of an afterglow is that of the expansion rate of the blast wave.
In the case of GRB~970508 an apparent  superluminal transverse expansion
was inferred from the evolution of the modulation index of the scintillating
flux at 8.5~GHz \citep{fkn97}.
Early in the evolution, the radio flux 
showed significant fluctuations (up to $\sim 50$\%), as would be expected 
due to interstellar
scintillation of a source of very small angular size.  This scintillation
gradually decreased with time, and became nearly imperceptible after 
$\sim 50$~days.
This evolution can be attributed to an increase of the angular size of the 
source with time.  The expansion rate derived from these observations
was $\sim 3\,{\mu}\mbox{\rm as}$ in $\sim 2$~weeks, which, at the redshift 
of the source (z=0.835), amounted to
a transverse expansion speed of $\sim 4$~times the speed of light.  Using 
the standard interpretation of superluminal motion, this would suggest
that the average bulk Lorentz factor of the blast wave $\sim 2$~weeks 
after the burst was  $\sim 4$ \citep{fkn97}.

In the case of GRB~030329 it has been possible to directly measure the
angular extent of the expanding source using Very Long Baseline 
Interferometry (VLBI) at several epochs over nearly 3 years following 
the burst 
\citep{tfb04, ptg07}.
These measurements show an apparent superluminal transverse expansion
rate in the early phase ($v \sim 6c$ at $\sim 20$ days after burst), which 
gradually becomes sub-luminal around
$\sim 1$~yr after the burst.  The evolution of the 
apparent transverse size can be used to distinguish between several 
possible models of post jet-break lateral expansion of the blast wave -- 
the available measurements on GRB~030329, however, are not strongly
constraining in this regard \citep{granot05, ptg07}.
The non-relativistic transition time derived from the VLBI measurements 
of GRB~030329 appear to be a factor of $\sim 2$ larger than that required 
to successfully model the  multi-wavelength light curves of the afterglow
\citep{ptg07, horst07}.
The reason for this discrepancy is yet to be fully understood.

   \subsection{Polarization}

One direct consequence of synchrotron emission is that the emission
from an individual particle is polarized. Due to the probably
random nature of the post-shock magnetic fields, the polarization
is likely to be averaged out and only a small degree will be left. The time at which linear polarization
is detectable is thought to be around the jet-break time.
Several (differing) models have been proposed, in which a collimated
jet and an off-axis line of sight conspire to produce an asymmetry which
leads to net polarization including one or several 90\degs\ changes
of the polarization angle \citep{ghi99, sari99}. This behaviour could provide
independent evidence for the jet structure of the relativistic
outflow.

\begin{figure}
\includegraphics[angle=270,width=\textwidth]{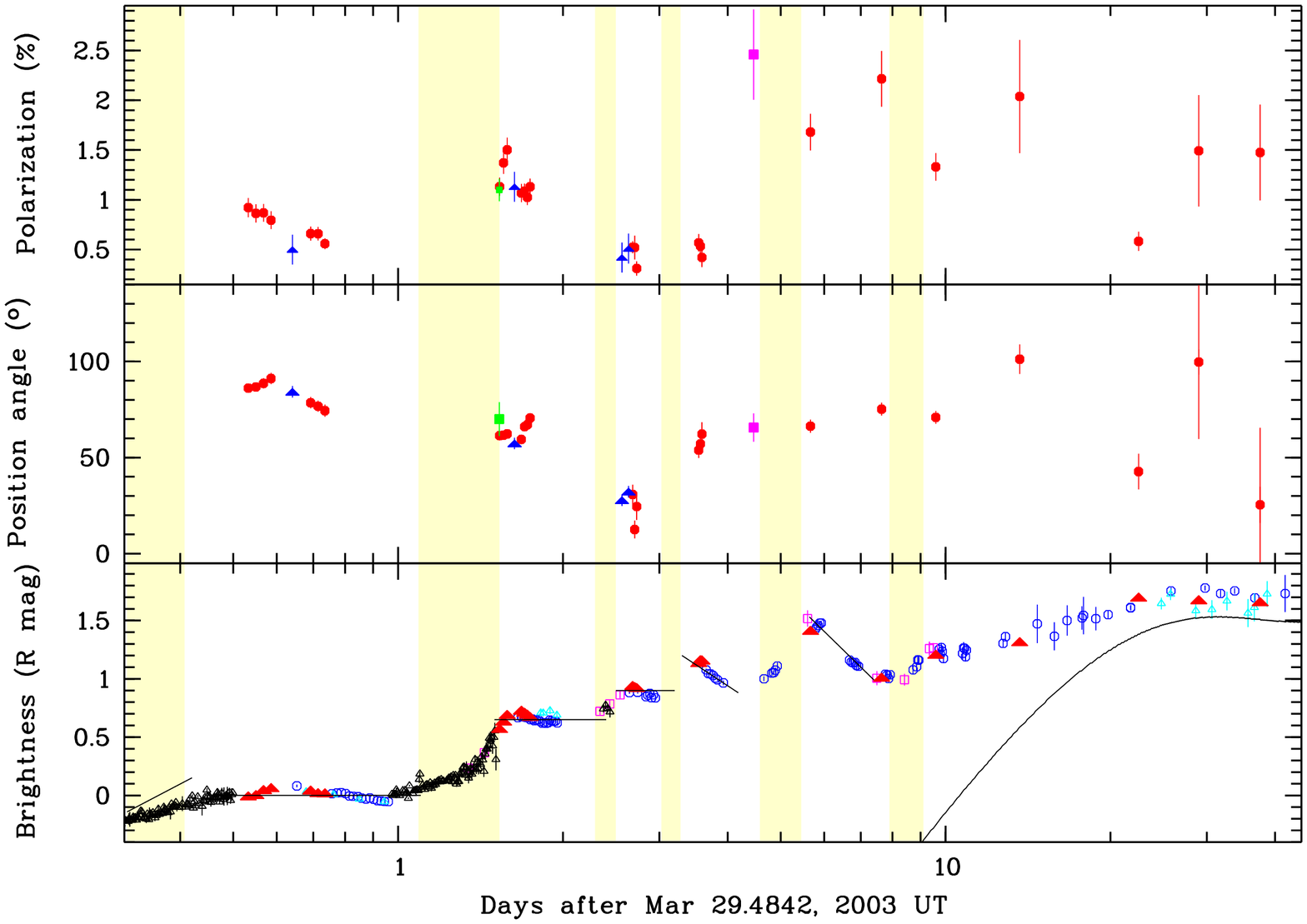}
\caption{Evolution of the polarization of the afterglow of GRB\,030329 during the first 38 days. The top and middle panels show the polarization degree in percent and the
position angle in degrees. 
The bottom panel shows the
residual $R$ band light curve after subtraction of a
power-law t$^{-1.64}$ describing the undisturbed decay during the time
interval $0.5-1.2$ days after the GRB,
thus leading to a horizontal curve. Gray bars mark re-brightening
transitions. Contributions from an underlying supernova (solid curved
line) do not become significant until $\sim$10 days after the GRB.
From \citet{gkr03} \label{030329pola}}.
\end{figure}

The observed polarization at optical wavelengths at later times is less than 
3\% \citep{hjo99, wij99, rol00} with 
one, debated, exception of 10\%
\citep{bers03}.
Because of these low-levels and the rapid decline of
the afterglow brightness during the first day, it has been
difficult to observe changes in the polarization as predicted by theory.
The by far most extensive observations of a light curve with fast variability
in polarization degree and angle (Fig. \ref{030329pola}) 
have been obtained for the afterglow of
GRB\,030329 \citep{gkr03}. This variability pattern does not follow
any of the model predictions, and is also not correlated with brightness.
The global behaviour is consistent with the interpretation
that the GRB is emitted in a relativistic jet with an initial opening
angle of 3\degs.
However, in this GRB afterglow several re-brightenings superposed to a power-law
decline have likely caused deviations from a
simple single-jet model, thus making it difficult to interpret.
The low level of polarization implies
that the components of the magnetic field parallel and perpendicular
to the shock do not differ by more than $\sim$10\,\%, and suggests an
entangled magnetic field, probably amplified by turbulence behind
shocks, rather than a pre-existing field.

Very recently, an optical polarization measurement of GRB\,090102
was achieved
at a time when the reverse shock emission was dominating the light curve \citep{sms10}.
The method uses a rotating polaroid which allows simultaneous measurements of 
the polarization degree of neighbouring stars but not the angle.
This implies that the constant polarization of the Galactic foreground ISM
could not be subtracted, and thus the measured  polarization of 10.2$\pm$1.3\%,
is likely an upper limit.
This relatively high level has been 
interpreted as evidence for the presence of large-scale ordered magnetic 
fields in the relativistic outflow. In the present case, the magnetisation,
i.e., the ratio of magnetic to kinetic energy, must have been fine-tuned
to near 1. Any value substantially larger than 1 would suppress the observed
reverse shock, while values well below 1 would not produce a net
polarization at the measured level.

\begin{figure}
\hspace{-0.5cm}\includegraphics[angle=0,width=7.5cm]{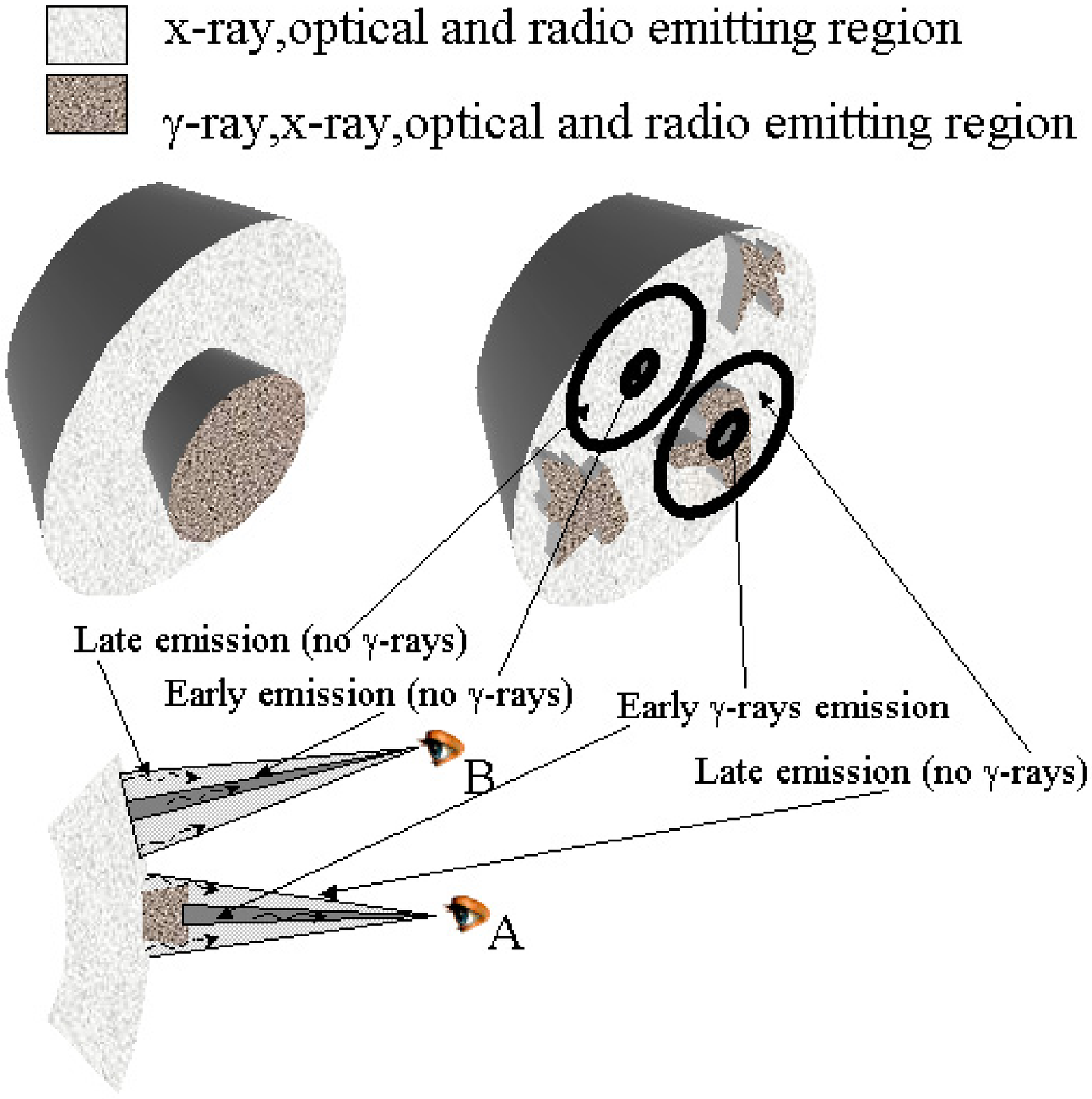}\hspace{-2.cm}
\includegraphics[angle=0,width=7.cm]{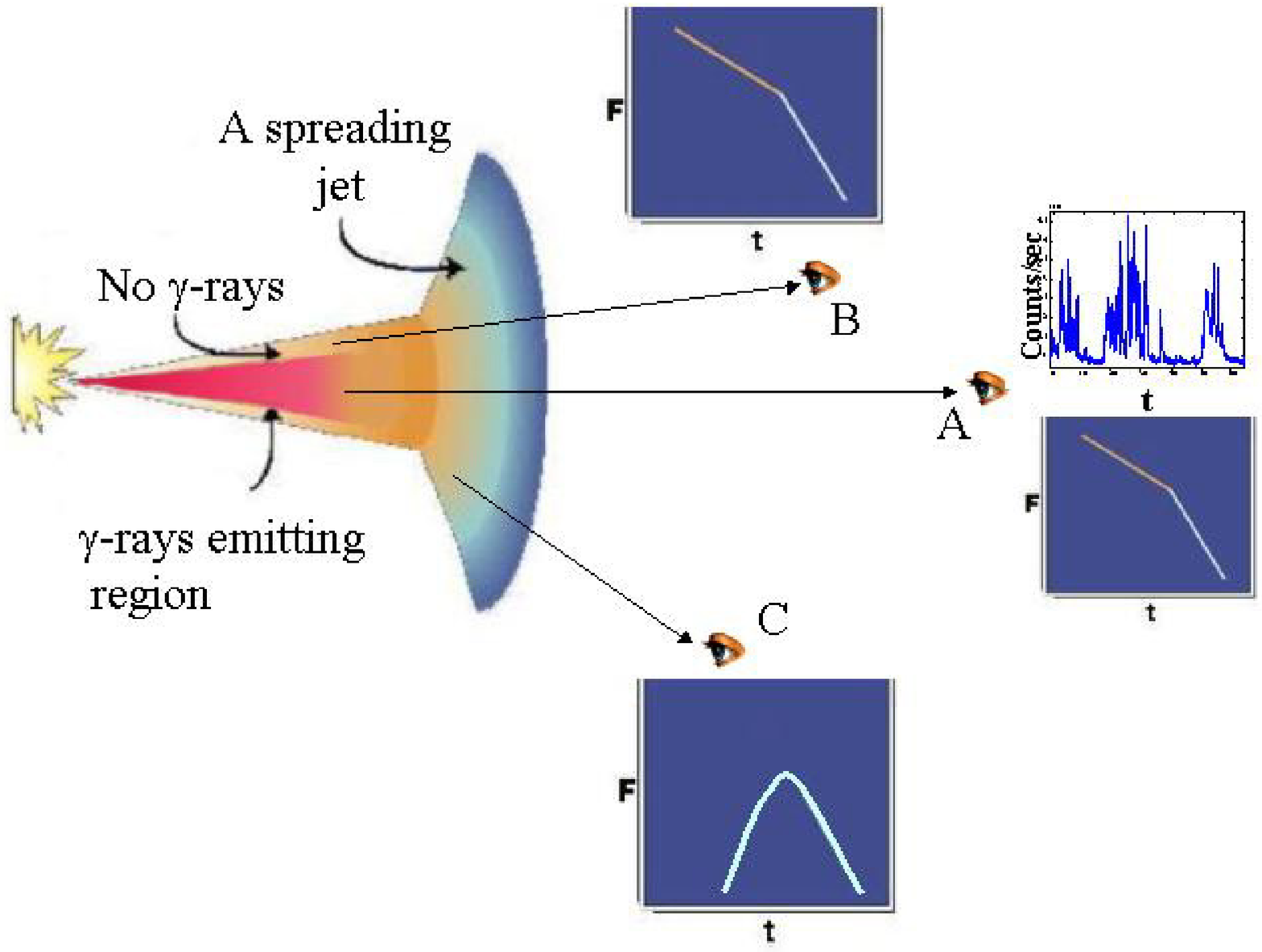}
\caption{{\bf Left:} Schematics of an on-axis orphan afterglow: prompt 
  gamma-rays are emitted only by some regions which can have either a 
  regular (upper left; cross-section in the lower picture) or 
   irregular structure (upper right). The ellipses
  describe the area seen by an observer at a given time. Observer A detects
  the early emission from a small region within the gamma-ray emitting region,
  and later an afterglow from a much larger region (regular GRB and afterglow).
  Observer B does not detect any gamma-rays, but detects a regular 
  (on-axis orphan) afterglow.
  {\bf Right:} An off-axis orphan afterglow is seen by observers which are 
   not within the initial relativistic jet. This emission is seen only after 
   the jet break. Observer A detects both, the GRB and the afterglow; 
   observer B detects the same afterglow but no gamma-rays, and observer C
   detects an off-axis orphan afterglow.
   From \citet{nap03} \label{orph}}.
\end{figure}


   \subsection{Orphan afterglows}

An exciting consequence of beaming is that there should exist GRBs
which develop a less beamed X-ray, optical, or radio afterglow, 
but for which we miss the
prompt GRB emission - the so-called orphan (Fig. \ref{orph})
afterglow (for a discussion see, e.g.,  
\citet{rhoad97, mes98, per98}).
Archival X-ray data have been  searched for such events, but none was found 
\citep{grind99, ghv00}.
In the optical, a small number of dedicated surveys was performed; there no candidate event was found in 125 hrs of monitoring
of a field of 256 sq. deg. with {\it ROTSE-I} to a limiting magnitude
of 15.7 \citep{kab02}.
\citet{vlw02}
searched for color-selected transients within 1500 sq. deg. of the
Sloan Digital Sky Survey ({\it SDSS}) down to $R = 19$ and found
only one unusual transient which was later identified as a radio-loud
AGN exhibiting strong variability \citep{gof02}.
A couple of interesting optical
transients were found in the $B$, $V$ and $R$-band Deep Lens Survey ({\it DLS})
transient search, within an area of 0.01 deg$^{-2}$ yr$^{-1}$ with a limiting
magnitude of 24. None of these could be positively associated
with a GRB afterglow \citep{bwb04} and all were later shown to have
been flares from M dwarfs in our Galaxy \citep{kur06}. In another unsuccessful
search using the {\it ROTSE-III} telescope array \citep{raa05} 
placed an upper limit on the rate of fading optical transients
with quiescent counterparts dimmer than $\sim$20th magnitude of
less than 1.9 deg$^{-2}$ yr$^{-1}$.
Finally, a monitoring project of $\sim$12 sq. deg. in 25 nights (at a typical spacing of 2 nights) down to a limiting magnitude of $R \sim 23$ mag found no afterglow
candidate, providing a  limit on the
collimation factor (ratio of the true rate of on-axis optical afterglows
to long-duration GRBs which produce observable optical afterglows)
of $<$12\,500 \citep{rgs06}.

In the radio band, orphan afterglows have been searched for by
combining the Faint Images of the Radio Sky at Twenty-centimeters ({\it FIRST})
and the NRAO VLA Sky Survey ({\it NVSS}), with the result of finding
9 afterglow candidates and implying a limit on the 
beaming factor of $f_b^{-1} \equiv (\theta^2/2) >13$
if all candidates, and $f_b^{-1} >90$ if none are associated with GRBs, 
respectively \citep{low02}. These authors also noted the, at first glance
anti-intuitive, fact that the number of orphan radio afterglows is smaller for
smaller jet opening angles in a flux-limited survey (for narrower beams 
each GRB has a lower energy and, therefore, is more difficult to  detect) 
Later, \citet{gay06} concluded that none of the transient objects was an 
orphan afterglow and set an upper limit for the beaming factor, 
$f_b^{-1} > 62$.
Recently, there is evidence for an orphan radio afterglow found in the
search for type Ibc SNe, through
the discovery of luminous radio emission from the seemingly ordinary 
type Ibc SN\,2009bb, which, however, requires a substantially relativistic 
outflow powered by a central engine \citep{scp10}. A mildly relativistic 
outflow was also observed in SN\,2007gr \citep{paragi10}. These detections 
indicate that, most likely, the relativistic energy content of Ibc SNe 
varies dramatically, while their total explosion energy maybe more standard. 

\section{Constraints from multi-wavelength afterglow observations}

   \subsection{Fireball parameters} 

\begin{figure}[bh]
\includegraphics[angle=270, width=7.cm]{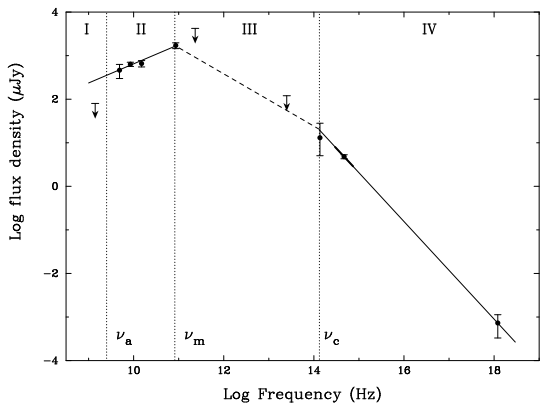}
\hfill\parbox[t]{4.cm}{\vspace*{1.0cm}\caption[SED970508]{X-ray-to-radio
   spectral energy distribution of  GRB\,970508 
    at 12.1 days after the burst. Indicated are the inferred values of the
   break frequencies $\nu_a$, $\nu_m$ and $\nu_c$. From \citet{gwb98}.
 \label{SED970508}}}
\end{figure}

The evolution of the blast wave in the fireball model is governed by
the total energy in the shock, the geometry of the outflow, and the
density structure of the ISM into which it is expanding 
(see also Chapters 7 and 8). The time
dependence of the radiated emission depends on the hydrodynamic evolution
and the distribution of energy between electrons and magnetic field 
\citep{sap99}. Unfortunately, only for few bursts sufficient data have
been collected in order to derive the fundamental physical parameters:
GRBs\,970508 (Fig. \ref{SED970508}) \citep{gwb98, wig99}, 
980329 \citep{yfh02}, 980703 \citep{fyb03} and 051111 \citep{blp06}.

Despite this sparse number of GRBs sampled, it is obvious
that the diversity in physical parameters is large: the ISM density
ranges between 0.1--500 cm$^{-3}$, total energies are 
10$^{51}$ to 10$^{53}$ erg, and the energy distribution between electrons
and magnetic field is consistent with equipartition.
Future observations are clearly warranted to improve our understanding
of the distributions in these parameters, and to what extent more
sophisticated models with more parameters are needed.

   \subsection{Environment}       

\subsubsection{Extinction}

Besides deriving the fireball parameters from spectral energy distributions
(SED), emphasis has also been given to the curvature of broad-band spectra
in the optical/near-infrared (NIR) region due to dust extinction, and in the
soft X-ray band due to absorption by gas. 

Effective neutral hydrogen absorption in excess of the Galactic foreground
absorption has taken a long way to get detected significantly in GRB
afterglow spectra. Originally not detected at all in the full sample
of BeppoSAX bursts \citep{ppp03}, a re-analysis of the brightest
13 X-ray afterglows  revealed statistically significant absorption in excess
of the Galactic one for two bursts \citep{sfa04}. Already 8 bursts of 17
observed with {\it Chandra} or XMM-{\it Newton} until Oct. 2004 show excess
absorption \citep{gcp06}. In the Swift era, excess absorption is detected
in the majority of bursts, in selected samples up to 85\% \citep{gkk10}.

In the optical/NIR, extinction measurements for a long time have been 
hampered by the lack of proper SED measurements, and the interrelation
of spectral slope, redshift and extinction. Early attempts therefore 
concentrated on deep NIR observations \citep[e.g.][]{khg03}.
In a first systematic way
\citep{kkz06} collected photometry of 19 bursts from the literature,
constructed light curves, shifted measurements of different filters to
a common epoch according to the light curve, and derived spectral slope
and extinction $A_{\rm V}$. While little evidence was found for substantial
$A_{\rm V}$, the prevalence of a SMC-like dust extinction curve was noted.
In the Swift era, UVOT observations provided more accurate $A_{\rm V}$
measurements, but for a sample which is strongly biased towards bright
and small-$A_{\rm V}$ bursts \citep{smp07, spo10}. Recently, the systematic
GRB follow-up with the P60 \citep{cfm06} and GROND instruments \citep{gbc08} 
provided the first unbiased view on the extinction properties 
(see section \ref{darksec}),
with a substantially larger fraction of bursts with moderate $A_{\rm V}$
\citep{ckh09, gkk10}.

\subsubsection{Wind vs. constant density profile}


The likely progenitor of long-duration GRBs is the stripped core of a 
massive star of initial mass \gax25 \msun, similar to a Wolf-Rayet star. 
The winds from these stars in our Galaxy have velocities of 1000--2500 km/s
and mass-loss rates of $10^{-5} - 10^{-4}$ \msun/yr. Before exploding
and creating a GRB, a Wolf-Rayet star is thus expected to be surrounded by
a medium with density $\rho \propto r^{-s}$, where $r$ is the distance from the
star, and $s$=2 for a stellar wind density profile and $s$=0 for a constant
interstellar medium (ISM) density.

The emission for a thin shell model expanding into a pre-blown wind 
has been calculated by \cite{chl99, chl00}, 
while that interacting with a constant ISM density ($s$=0) can be found in
\cite{wax97, spn98}. The appearance of the spectrum (as determined by the
power law index $p$ of the electron distribution) at a given time
is similar for both cases, but the evolution is different. At high frequency,
e.g., optical/X-rays, for $s$=0 the flux evolution goes from adiabatic 
($\propto t^{-(3p-3)4}$) to cooling ($\propto t^{-(3p-2)4}$) while
for $s$=2 it goes from cooling ($\propto t^{-(3p-2)4}$) to adiabatic
($\propto t^{-(3p-1)4}$). While cooling, the two cases have the same
spectrum and decline. At low frequency (radio), the flux evolution is
$\propto t^{1/2}$ for $s$=0, but can make a transition from  $\propto t$
to constant for $s$=2 \citep{chl99}.

Although wind models are indicated for some observed afterglows,
the majority are better described by constant density environments.


   \subsection{Progenitors} 

\subsubsection{Long-duration GRBs}

Based on the observed supernova connection \citep{wob06}, 
the progenitors of long-duration 
GRBs are intimately connected to supernovae. These progenitors must have lost
their hydrogen envelope prior to the supernova explosion. In order to
explain the observed statistics, the progenitors must be massive and
frequent enough:
A comparison of the supernova features in GRB afterglow light curves
with those of non-GRB related stripped-envelope supernovae shows that
the GRB-SN have, on average, considerably higher kinetic energies
and ejected masses \citep{rich09}.
Which special circumstances lead to the final occurrence
of a GRB is not fully understood.
There are certain mass ranges which make an explosion more difficult,
but this depends on the rotation of the progenitor \citep{fry99, whw02}. 
Also, besides single 
star channels also binary channels have been proposed 
\citep[e.g.][]{svr02, pmn04, frh05},
making a specific prediction difficult. 

The role of rotation in supernovae is a long-standing question going
back to \cite{hoy46}, but for GRB-SN there is general consensus that
rotation is required. However, the details \citep{hlw00, spr02}
as well as the questions of binarity \citep{yol05} remain open.
The rotation as well as the mass of the GRB progenitor are crucially 
influenced by mass loss. Replenishment of material lost from the
surface will reduce the rotation rate, and mass loss will make the 
star lighter at the time of explosion.

Preferred GRB scenarios thus have a small mass-loss rate, particularly
in the Wolf-Rayet phase, at which the mass loss rate is smaller for
low metallicity \citep{vdk05}. This has led to the general
expectation that GRBs should favour low metallicity regions \citep{mfw99}.

\begin{figure}[ht]
\includegraphics[angle=270, width=9.2cm]{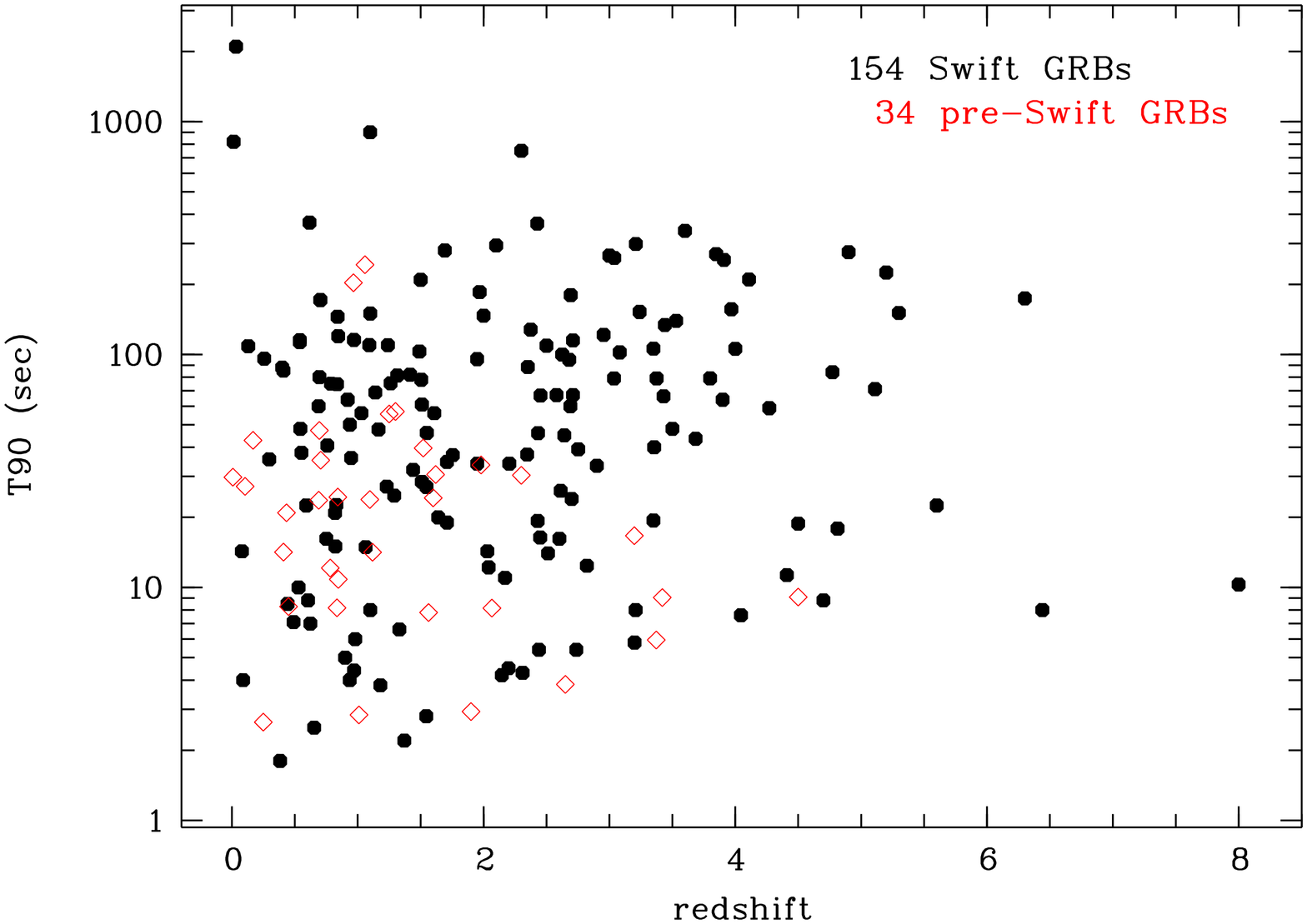}
  \hfill\parbox[t]{3.2cm}{\vspace*{-0.1cm}\caption[zvsT90]{Rest-frame
    duration of long GRBs
    versus redshift. The predicted trend of larger $T_{90}$ with
    redshift $z$, is not obvious. Filled symbols denote {\it Swift} GRBs,
   while open triangles denote pre-{\it Swift} era bursts (which have a 
   different bias in the $T_{90}$ determination).
 \label{zvsT90}}}
\end{figure}

The above line of thoughts might suggest that high-redshift bursts should have,
in general, longer duration than nearby bursts. Lower metallicity 
in the early phases of the Universe would leave more mass and rotation energy
with the progenitor due to less strong winds, which in turn
should have a consequence on the time scale of accretion and/or
fall-back. With the present sample of GRBs with redshift no such
correlation is seen (Fig. \ref{zvsT90}), indicating that the duration 
measure $T_{90}$ does not (only) depend on mass and rotation.

An interesting point, however, is that a fraction of $\sim$8\% of 
long-duration bursts have rest-frame durations \lax 1 sec (independent of
redshift; see Fig. \ref{zvsT90}). This poses the question of
how massive stars can produce burst of such short duration? 
Since the fall-back of material from the envelope is of order 100 sec,
this high rate of intrinsically short bursts related to massive stars
may imply that the burst duration is rather determined by the
ejection of the jet or the dissipation of the kinetic energy of the jet.

Population synthesis models show that for the redshift range 6--10
the majority of GRB progenitors are Population II stars \citep{bhf10},
as Population III (metal free) stars have already finished their evolution
and Population I (metal rich) stars are just beginning to form.
The peak of the long-duration GRB rate depends on the poorly constrained
metallicity evolution and peaks at $z \sim 7$ (3) for efficient/fast 
(inefficient/slow) mixing of metals \citep{bhf10}.

\subsubsection{Short-duration GRBs}

The association of short GRBs with early-type galaxies 
\citep{gsb05, bpp06}, and the burst 
localizations being relatively distant from the center of the host,
have supported the earlier conjecture that the progenitors of short
GRBs are related to an old stellar population,
namely binary systems composed of two compact objects that merge
after their orbit has decayed through gravitational wave emission
\citep{elp89}.

Interestingly, however, some short GRBs are associated with small, 
star-forming galaxies, and explode close to their center \citep{tko08}.
In a recent census actually 4$\times$ more bursts reside in
star-forming galaxies than in elliptical galaxies \citep{ber09}. This 
association spurred discussion on different families of progenitors for short
GRBs, including different compact object types \citep{bpb06},
proto-magnetars \citep{mqt08},
or a tighter connection to the star-formation evolution similar to
long-duration bursts \citep{vzo10}.
A comparison of the luminosities, star formation rates and metallicities
of a sample of hosts of short and of long-duration bursts shows,
however, that short burst hosts appear to be drawn uniformly from the
underlying field galaxy distribution. This suggests a wide age distribution
of several Gyr for the progenitors of short GRBs \citep{ber09}, though
this is also consistent with the possibility that the associations
of short GRBs to host galaxies are systematically flawed.

   \subsection{Jet opening angle}      

As described above, orphan afterglow searches have not yet been
sensitive enough to constrain the beaming fraction in a sensible way.
Similarly, polarisation measurements have not (yet) confirmed
that breaks seen in light curves are jet breaks. Thus, estimates
of jet opening angles rely exclusively on the identification
of observed breaks in the afterglow light curves, and their association
with a jet break. In the HETE-II and BeppoSAX era, the requirement
of achromaticity was only loosely applied, due to the sparse coverage
of radio, optical/NIR and/or X-ray measurements \citep{fks01, bfk03}. 
These first attempts found the surprising result that the seemingly
most energetic bursts also had the smallest beaming factor, so that
the true, beaming corrected energy release was strongly clustered.

In the Swift era, 
measuring the jet opening angle has been a more challenging task:
the much better database of X-ray and 
optical/NIR follow-up of Swift bursts has made identifying achromatic breaks
much more rare \citep{rlb09}. In the few clear cases, the distribution of 
jet break times ranges from a few hours to a few weeks with a median of 
$\sim$1 day \citep{rlb09}, implying opening angles of few to about 20\degs.
Another uncertainty, which already plagued the first attempts,
is the problem of constant ISM or wind density profile, leading to opening
angle estimates differing by up to a factor of 2.


With knowledge of the jet opening angle, an estimate of the
true rate of GRBs can also be made. In the Universe, there are about
5 supernovae per second \citep{mdp98}. The exposure and sky-coverage rate
corrected GRB rate is about 3 per day. Correcting this for a mean
beaming factor of 300 implies that throughout the Universe, the GRB rate 
is only about 0.2\% of the SN rate, and thus a rare phenomenon among
core-collapse SNe. If the GRB rate is strongly dependent on metallicity,
this fraction will be higher at large redshift.

   \subsection{Distance and Energetics}       

\begin{figure}[ht]
 \includegraphics[angle=270, width=8.8cm]{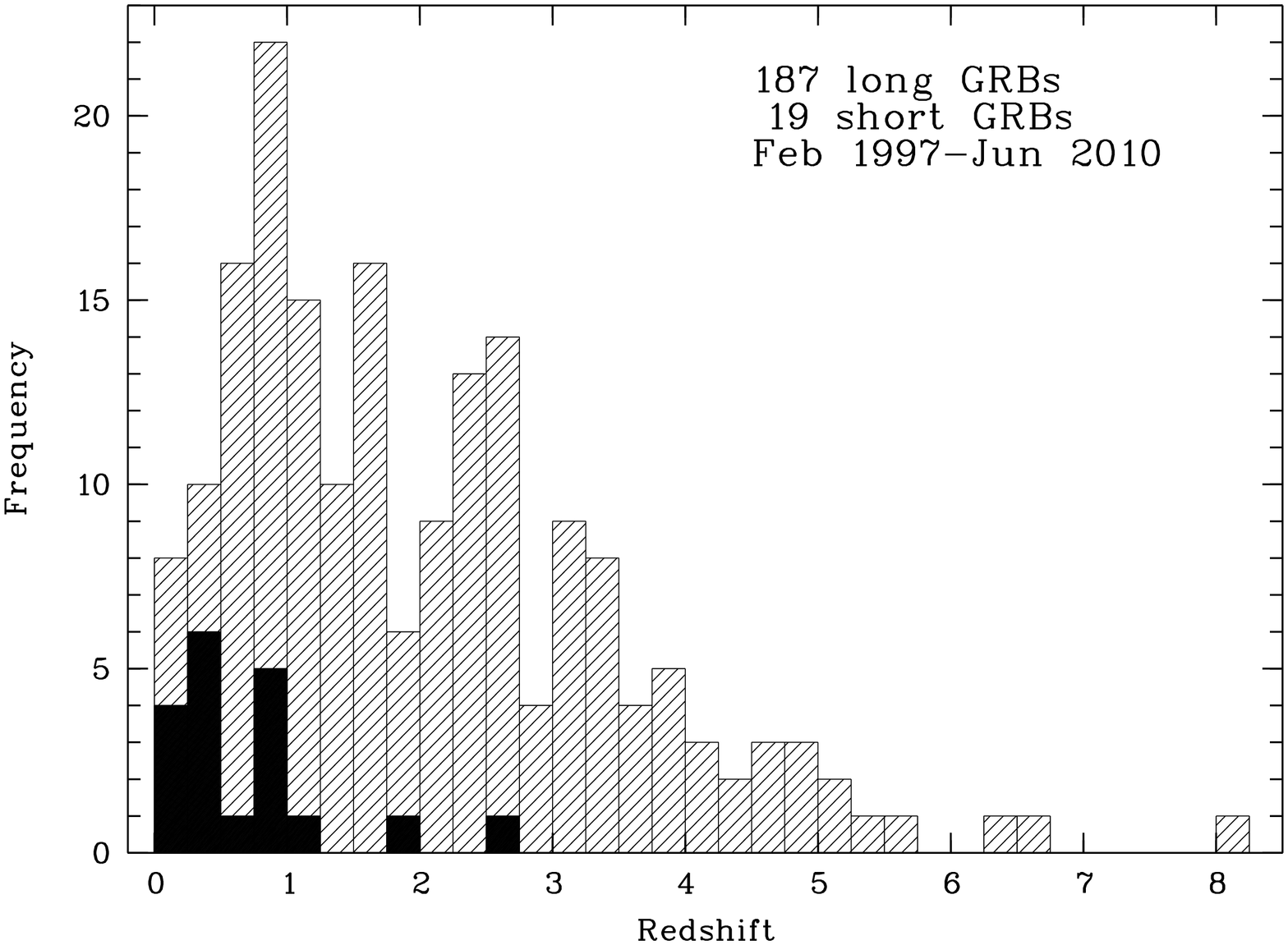}
 \hfill\parbox[t]{3.5cm}{\vspace*{3.5cm}\caption[zdistr]{Observed redshift 
     distribution of long- (grey) and 
     short-duration (black) GRBs as of June 2010. 
 \label{z-dis}}}
\end{figure}

Beyond the prompt emission fluence, two observables are required to 
determine the energetics of gamma-ray bursts: their distance,
and their jet opening angles. Since the discovery of afterglows, redshifts
have been measured for nearly 200 bursts (Fig. \ref{z-dis}); 
their isotropic equivalent energy is in the range of 10$^{51}$ to $10^{54}$ erg.
However, with only a few jet opening angles
measured, the distribution of beaming-corrected energetics remains
poorly constrained.

For a handful of bursts detected recently with {\it Fermi} in the
0.1--several GeV range, light curve breaks or limits could be derived, and
beaming-corrected energies determined. Four of these bursts, namely
GRB\,080916C, 090902B, 090926A and 090323, have beaming-corrected energies 
$E_{\gamma}$ of $>$2-5 $\times$ 10$^{52}$ erg \citep{gck09, mkr10,cfh10b,rsk10}, 
among the highest ever measured.
Interestingly, their jet opening angles are not particularly narrow.
Values in excess of 10$^{52}$ erg have been reported
for another {\it Fermi}/Large Area Telescope (LAT) detected burst \citep{gck09}
and a number of {\it Swift} bursts 
\citep{cfh10a}. These results indicate that the distribution of $E_{\gamma}$ 
is broad (at least a factor of 30) and not compatible with a standard
candle \citep{fks01,  bfk03}. Furthermore,
while being compatible with the Amati ($E_{\rm peak} - E_{\rm iso}$) relation 
at the 2$\sigma$ level \citep{afg09},
these very luminous GRBs with high values of $E_{\rm peak}$ are not compatible 
with the $E_{\rm peak} -  E_{\gamma}$ relationship
\citep{ggl04}. Both these correlations are heuristic, based on prompt
gamma-ray emission properties, and have survived over the last decade with
measurements by various instruments. $E_{\gamma}$ is the beaming-corrected
version of $E_{\rm iso}$, the total bolometric energy released by a burst
(see chapter {\bf ???}).

GRBs being beamed, not only the energy per burst is reduced by 2-3 orders 
of magnitude, but also their frequency is increased correspondingly since an 
observer will miss most of the narrow-beamed events. This in turn has 
implications on the GRB rate, and its relation to the star-formation rate.

   \subsection{Cosmology}

GRB afterglows are bright enough to be used
as pathfinders into the very early universe, independent of
whether or not the GRB and/or afterglow phenomenon is fully understood.

In contrast to stationary sources at high redshift,
GRB afterglows do not appear
substantially fainter at increasing $z$. Relativistic time dilation
implies that the observations of GRBs at the same time $\Delta$t after
the GRB event in the observers frame ({\em on Earth}) will be observed
at different times in the source frame, e.g. at {\em earlier} times for
more distant GRB. At this {\em earlier} time the GRB is intrinsically
brighter, thus partly compensating the larger distance.

While it seems unlikely that GRBs will soon be used to derive an
accurate Hubble-diagram
and to constrain cosmological parameters below the accuracy provided
by other methods, there are a few
other implications of high-$z$ GRB studies for cosmology:
\begin{itemize}
\item Since long-duration GRBs are related to the death of massive
stars, it is likely that high-$z$ GRBs exist, as examplified by the
recent discoveries of GRBs at redshift 6.7 \citep{gkf09}
and 8.2 \citep{tfl09, sdc09}. Theoretical
predictions range between a few up to 50\% of all GRBs being at $z>5$
\citep{msc00, lar01, bromm03}, while observations indicate a level of 
5\% (see also Chapter 14).
With WMAP data and theoretical expectations pointing towards the first 
star-formation occurring at $z \sim 20-30$ \citep{kogu03},
further redshift records can be expected in the near future.
Hopefully, also the spectroscopic follow-up will be improved, thus allowing us
to use these high-z bursts to expand our understanding of the early Universe
with respect to metallicity evolution or re-ionization history.
\item WMAP data also suggest that the onset of re-ionisation happened
at $z=11-20$ \citep{kogu03}. Because WMAP
only provides an integral constraint on the re-ionisation
history of the universe, it has led to the speculation that
re-ionisation was either an extended process or happened more than once.
Since the intrinsic luminosity as well as
the number density of quasars are expected to fade rapidly beyond $z \sim 6$,
only GRBs are suitable to be used as bright beacons to illuminate
the end of the dark age \citep{bar01, loeb01, mir03},
and potentially allow us to probe the re-ionisation history of the early
Universe \citep{ino03}.
\item Extensive monitoring of afterglows would help to constrain their
local environment, and could allow us to tell whether GRB afterglows
are decelerated by the intergalactic medium with an increasingly higher density at higher
redshift, or by a stratified constant density medium in a bubble cleared
by the progenitor star \citep{gou03}.
\item Studying the distribution and absorption line properties
of GRB host galaxies would shed light onto the cosmological structure
formation and star forming history \citep{mao98}.
\end{itemize}

\section{Prospects for the future}

With the launch of Fermi, the GRB field has entered a new era as
emphasis is re-directed again to the main emission mechanism.
Yet, there are at least two aspects which relate to the afterglow
phenomenon: First, the origin of the delayed GeV emission has
been proposed to be afterglow emission \citep[e.g.][]{ggn10}.
Second, it turns out that the very energetic Fermi/LAT bursts are
also those with particularly large beaming corrected energies 
\citep{mkr10, cfh10b}.

One might also hope that if GBM positions  could be freed of their
systematic errors of 5--12 degrees \citep{bcm09}, the recovery of
optical afterglows of bright GBM bursts may have a large impact on the
question of jet breaks, and consequently on the beaming angle
distribution.

Current and near-future improvements in our ground-based facilities
include 
\begin{itemize} \vspace{-0.22cm}
\item  the routine use of a spectrograph with a wide wavelength coverage
  from the atmosphere cut-off to the near-infrared (X-Shooter) at
  the ESO/VLT, allowing the detection of absorption and emission lines
  over the full wavelength range accessible from ground. 
\item the upgrade of the VLA to substantially higher sensitivity
  (EVLA), and the starting operation of LOFAR and ALMA, the latter covering
  the peak of the synchrotron spectrum of GRB afterglows, allowing
  a substantial fraction of, if not all, afterglows to be detected
  and calorimetry to be used to determine the GRB energetics;
\item the upgrades of air Cerenkov telescopes to lower energy thresholds
  will allow us to cover a larger distance range before photon-photon
  interactions attenuate the signal.
\vspace{-0.22cm}
\end{itemize}
These instruments will change the number of afterglow discoveries
and the amount of data per afterglow dramatically, thus allowing
completely new studies to be performed.

In the field of non-electromagnetic signatures, both neutrino and gravitational
wave detectors are getting close to the expected fluxes from GRBs.
IceCube \citep{aaa10} and ANTARES \citep{bou10} should soon be able to 
detect the
typically 1-10 GeV neutrinos which are expected to be produced in the 
shocks related to GRBs, or even inelastic proton-neutron collisions
in shock-free environments, and thus would confirm that protons are accelerated.
Similarly, the Advanced LIGO interferometer,
once coming online around 2014, should detect of order 10 
neutron star mergers related to short GRBs, 
up to distances of 200 Mpc \citep{gul09}, and could provide insight
into the inner engine.

\bigskip

{\bf Acknowledgements:} 
I thank Dipankar Bhattacharya for writing 
section \ref{dipank} and Evert Rol for discussions during the early stage 
of this review.
I acknowledge S. Klose and A. Rau for comments on an
earlier version of this manuscript, 
and D.A. Kann for help in preparing Tab. \ref{brightAG} as well as
a proof reading of the manuscript.

\bigskip





\begin{thereferences}{99}


\bibitem[Abbasi et al.(2010)]{aaa10} Abbasi R., Abdou Y., Abu-Zayyad T.
   et al. (2010). \textit{ApJ} \textbf{710}, 346.

\bibitem[Akerlof et al.(1999)]{abb99} Akerlof C., Balsano R.,
Barthelmy S.,  \etal . (1999). \textit{  Nat.} \textbf{398}, 400.

\bibitem[Amati et al.(2009)]{afg09} Amati L., Frontera F., Guidorzi C. (2009). \textit{
 A\&A} \textbf{508}, 173. 

\bibitem[Band \& Hartmann(1992)]{bah92}
    Band, D. L.  \& Hartmann, D.H. (1992). \textit{ ApJ} \textbf{386}, 299. 

\bibitem[Barkana \& Loeb(2001)]{bar01} Barkana R., Loeb A. (2001). \textit{ 
  Phys. Rep.} \textbf{349}, 125. 

\bibitem[Barthelmy et al.(1996)]{bbc96} Barthelmy S.D., Butterworth P.S., 
  Cline T.L. \etal . (1996).  In: \textit{Gamma-Ray Bursts},
Eds. C. Kouveliotou \etal, AIP \textbf{384}, p. 580. 

\bibitem[Barthelmy et al.(2005)]{bcg05} Barthelmy S.D., Cannizzo, J.K., 
   Gehrels, N. et al. (2005).  \textit{ApJ} \textbf{635}, L133.

\bibitem[Becker et al.(2004)]{bwb04} Becker A.C., Wittman D.M., 
   Broeshaar P.C., et al. (2004). \textit{  ApJ} \textbf{611}, 418.

\bibitem[Belczynski et al.(2006)]{bpb06} Belczynski K., Perna R., Bulik T.
  et al. (2006). \textit{ApJ} \textbf{648}, 1110.

\bibitem[Belczynski et al.(2010)]{bhf10} Belczynski K., Holz D.E., Fryer C.L.
   et al. (2010). \textit{ApJ} \textbf{708}, 117.

\bibitem[Berger, Kulkarni \& Frail(2004)]{bkf04} Berger, E., 
   Kulkarni, S.R., Frail, D.A. (2004). \textit{ ApJ} \textbf{612}, 966. 

\bibitem[Berger et al.(2006)]{bpc06} Berger E., Penprase B.E., Cenko S.B. 
   et al. (2006). \textit{    ApJ} \textbf{642}, 979. 

\bibitem[Berger(2009)]{ber09} Berger, E. (2009). \textit{ ApJ} \textbf{690}, 231.

\bibitem[Bersier et al.(2003)]{bers03} Bersier, D., et al. (2003). \textit{
ApJ} \textbf{583}, L63.

\bibitem[Bj\"ornsson, Gudmundsson \& Johannesson(2004)]{bgj04} 
 Bj\"ornsson G., Gudmundsson E.H., Johannesson G. (2004). \textit{ ApJ} \textbf{615}, L77.

\bibitem[Blake et al.(2005)]{bbs05} Blake C.H., Bloom J.S., Starr D.L., et al.
  (2005). \textit{ Nat.} \textbf{435}, 181. 


\bibitem[Bloom et al.(2003)]{bfk03} Bloom J.S., Frail D.A., Kulkarni S.R.
  (2003). \textit{ ApJ} \textbf{594}, 674. 

\bibitem[Bloom et al.(2006)]{bpp06} Bloom J.S., Prochaska J.X., Pooley D. 
  et al. (2006). \textit{ ApJ} \textbf{638}, 354.

\bibitem[Bloom et al.(2009)]{bpl09} Bloom J.S., Perley D.A., Li W. et al. 
  (2009). \textit{ ApJ} \textbf{691}, 723.

\bibitem[Bo\"{e}r et al.(1988)]{bo88} Bo\"{e}r M., \etal (1988). \textit{ 
  A\&A} {\bf 202}, 117.

\bibitem[Bontekoe et al.(1995)]{bon95} Bontekoe T.J.R., Winkler C.,
 Stacy J.G., Jackson P.D. (1995). \textit{
Ap\&SS} \textbf{231}, 285.

\bibitem[Bouwhuis et al.(2010)]{bou10} Bouwhuis M. on behalf of the
ANTARES collaboration (2010). In Proc. of 31th ICRC, Lodz, arXiv:1002.0701.

\bibitem[Briggs et al.(2009)]{bcm09} Briggs M.S., Connaughton V., Meegan C.A.
   et al. (2009). AIP Conf. Proc. \textbf{1133}, p. 40.

\bibitem[Bromm \& Loeb(2002)]{bromm03} Bromm V., Loeb A. (2002). \textit{ 
 ApJ} \textbf{575},  111. 

\bibitem[Burrows et al.(2005)]{brf05} Burrows, D.N., Romano, P., Falcone, A.
    et al. (2005). \textit{Sci.} \textbf{309}, 1833.

\bibitem[Burrows et al.(2010)]{bur10} Burrows D.N. (2010).  in 
  {\it Deciphering the ancient Universe
  with GRBs}, Kyoto, Apr. 2010, AIP (in press)

\bibitem[Butler et al.(2006)]{blp06} Butler N.R., Li W., Perley D. et al. (2006). \textit{
    ApJ} \textbf{652}, 1390.

\bibitem[Cenko et al.(2006)]{cfm06} Cenko S.B., Fox D.B., Moon D.-S.
  et al. (2006). \textit{PASP} \textbf{118}, 1396. 

\bibitem[Cenko et al.(2009)]{ckh09} Cenko S.B., Keleman J., Harrison F.A.,
  et al. (2009). \textit{ ApJ} \textbf{693}, 1484. 

\bibitem[Cenko et al.(2010a)]{cfh10a} Cenko S.B., Frail D.A., Harrison F.A. 
  et al. (2010a). \textit{  ApJ} \textbf{711}, 641.

\bibitem[Cenko et al.(2010b)]{cfh10b} Cenko S.B., Frail D.A., Harrison F.A. 
  et al. (2010b). \textit{  ApJ} (subm.; arXiv:1004.2900).

\bibitem[Chen et al.(2005)]{cpb05} Chen H.-W., Prochaska J.X., Bloom J.S., 
    et al. (2005). \textit{ ApJ} \textbf{634}, L25.

\bibitem[Chevalier \& Li(1999)]{chl99} Chevalier R.A., Li Z.-Y. (1999). \textit{
   ApJ} \textbf{520}, L29.

\bibitem[Chincarini et al.(2010)]{cmm10} Chincarini G., Mao J., Margutti R. 
   et al. (2010). \textit{ MN} (subm., arXiv:1004.0901)

\bibitem[Chevalier \& Li(2000)]{chl00} Chevalier R.A., Li Z.-Y. (2000). \textit{
   ApJ} \textbf{536}, 195. 

\bibitem[Connaughton(2002)]{conn02} Connaughton V. (2002). \textit{ ApJ} \textbf{567}, 1028.

\bibitem[Costa et al.(1997)]{cos97} Costa E., Frontera F., Heise J. \etal (1997). \textit{
   Nat.} \textbf{387}, 783. 

\bibitem[Covino et al.(2008)]{cdk08} Covino S., D'Avanzo P., Klotz A.
 (2008). \textit{ MN} \textbf{388}, 347.

\bibitem[Curran et al.(2008)]{cso08} Curran P.A., Starling R.L.C., 
   O'Brien P.T. et al. (2008). \textit{ A\&A} \textbf{487}, 533.

\bibitem[Della Valle et al.(2006)]{dcp06} Della Valle M., Chincarini G., 
    Panagia N. et al. (2006). \textit{ Nat.} \textbf{444}, 1050.

\bibitem[Denisenko \& Terekhov(2008)]{det08} Denisenko D.V., Terekhov O.V. 
   (2008). \textit{Astron. Lett.} \textbf{34}, 298.

\bibitem[De Pasquale et al.(2003)]{ppp03} De Pasquale M., Piro L., Perna R.,
  et al. (2003). \textit{ApJ} \textbf{592}, 1018.

\bibitem[Dessauges-Zavadsky et a.(2006)]{dzcp06} Dessauges-Zavadsky M., 
  Chen H.-W., Prochaska J.X. et al. (2006). \textit{ApJ} \textbf{648}, L89.

\bibitem[Eichler et al.(1989)]{elp89} Eichler D., Livio M., Piran T.,
   Schramm D.N. (1989). \textit{Nat.} \textbf{340}, 126.

\bibitem[Falcone et al.(2006)]{fbl06} Falcone A.D., Burrows D.N., Lazzati D.
   et al. (2006). \textit{ ApJ} \textbf{641}, 1010.

\bibitem[Frail et al.(1997)]{fkn97} Frail, D.A., Kulkarni, S.R., Nicastro L. 
 et al. (1997). \textit{Nat.} \textbf{389}, 261.

\bibitem[Frail, Waxman \& Kulkarni(2000)]{fwk00} Frail, D.A., Waxman, E., 
Kulkarni, S.R. (2000). \textit{ ApJ} \textbf{537}, 191. 

\bibitem[Frail et al.(2001)]{fks01} Frail D.A., Kulkarni, S.R., Sari R., 
  et al. (2001). \textit{ ApJ} \textbf{562}, L55. 

\bibitem[Frail et al.(2003a)]{frail+03} Frail, D.A., Kulkarni, S.R., 
 Berger, E. et al. (2003a). \textit{  AJ} \textbf{125}, 2299.

\bibitem[Frail et al.(2003b)]{fyb03} Frail, D.A., Yost S.A., Berger E. et al.
   (2003). \textit{ ApJ} \textbf{590}, 992.

\bibitem[Frail et al.(2005)]{frail+05} Frail, D.A., Soderberg, A.M., 
  Kulkarni, S.R. et al (2005). \textit{ ApJ} \textbf{619}, 994.

\bibitem[Frontera et al.(1998)]{fga98} Frontera F., Greiner J., 
Antonelli L.A., et al. (1998). \textit{
A\&A} \textbf{334}, L69. 


\bibitem[Fryer(1999)]{fry99} Fryer C.L. (1999). \textit{ApJ} \textbf{522}, 413.

\bibitem[Fryer \& Heger(2005)]{frh05} Fryer C.L., Heger A. (2005).
     \textit{ApJ} \textbf{623}, 302.

\bibitem[Fynbo et al.(2001)]{fyn01} Fynbo J.P.U., Jensen B.L., Gorosabel J. 
  et al. (2001). \textit{ A\&A} \textbf{369} 373. 

\bibitem[Fynbo et al.(2006)]{fwt06} Fynbo J.P.U., Watson D., Th\"one C.C. 
  et al. (2006). \textit{  Nat.} \textbf{444}, 1047. 

\bibitem[Fynbo et al.(2009)]{fjp09} Fynbo J.P.U., Jakobsson P., Prochaska J.X.
   et al. (2009). \textit{ApJS} \textbf{185}, 526

\bibitem[Ga\-la\-ma et al.(1998a)]{gvp98} Galama T.J., Vreeswijk, P.M., 
  van Paradijs, J. et al. (1998a). \textit{Nat.} \textbf{395}, 670.

\bibitem[Galama et al.(1998b)]{gwb98} Galama T.J., Wijers R.A.M.J.,
   Bremer M. et al. (1998b). \textit{ ApJ} \textbf{500}, L97. 

\bibitem[Gal-Yam et al.(2002)]{gof02} Gal-Yam A., Ofek E.O., Filippenko A.V., 
  et al. (2002). \textit{ PASP} \textbf{114}, 587.

\bibitem[Gal-Yam et al.(2006a)]{gfp06} Gal-Yam A., Fox D.B., Price P.A. et al.
   (2006a). \textit{ Nat.} \textbf{444}, 1053.

\bibitem[Gal-Yam et al.(2006b)]{gay06} Gal-Yam A., Ofek E.O., Poznanski D.
  et al. (2006b). \textit{ ApJ} \textbf{639}, 331.

\bibitem[Gehrels et al.(2005)]{gsb05} Gehrels N., Sarazin C.L., O'Brien P.T. et al. (2005). \textit{
  Nat.} \textbf{437}, 851.

\bibitem[Gehrels et al.(2006)]{gnb06} Gehrels N., Norris J.P., Barthelmy S.D., et al. (2006). \textit{
   Nat.} \textbf{444}, 1044. 

\bibitem[Gehrels, Ramirez-Ruiz \& Fox(2009)]{grf09} Gehrels N., 
  Ramirez-Ruiz E., Fox D.B. (2009). \textit{ ARAA} \textbf{47}, 567. 

\bibitem[Gendre et al.(2006)]{gcp06}
Gendre B., Corsi A., Piro L. (2006). \textit{A\&A} \textbf{455}, 803

\bibitem[Gendre et al.(2010)]{gkp10} Gendre B., Klotz A., Palazzi E. et al. 
  (2010). \textit{ MN} (in press; arXiv:0909.1167)

\bibitem[Genet \& Granot(2009)]{geg09} Genet F., Granot J. (2009). \textit{ 
MN} \textbf{339}, 1328.

\bibitem[Ghirlanda, Ghisellin \& Nava(2010)]{ggn10} Ghirlanda G., Ghisellini G.,
 Nava L. (2010). \textit{A\&A} \textbf{510}, L7.

\bibitem[Ghisellini \& Lazzati(1999)]{ghi99} Ghisellini G., Lazzati D. (1999). \textit{
   MN} \textbf{309}, L7. 

\bibitem[Ghirlanda, Ghisellini \& Lazzati(2004)]{ggl04} Ghirlanda G., 
  Ghisellini G., Lazzati D. (2004). \textit{ ApJ} \textbf{616},  331. 

\bibitem[Goad et al.(2007)]{gpg07} Goad M.R., Page K.L., Godet O. et al. (2007). \textit{ 
   A\&A} \textbf{468}, 103.

\bibitem[Godet et al.(2007)]{gpo07} Godet O., Page K.L., Osborne J. et al. 
    (2007). \textit{ A\&A} \textbf{471}, 385.

\bibitem[Gou et al.(2004)]{gou03} Gou L.J., M\'{e}sz\'{a}ros P., Abel T., Zhang B. (2004). \textit{
ApJ} \textbf{604}, 508.

\bibitem[Granot, Ramirez-Ruiz \& Loeb(2005)]{granot05} Granot, J., 
Ramirez-Ruiz, E., Loeb, A. (2005). \textit{ ApJ}, \textbf{618}, 413.

\bibitem[Greiner et al.(1987)]{gfw87} Greiner J., Flohrer J., Wenzel W., 
  Lehmann T. (1987). \textit{  ASS} {\bf 138}, 155.

\bibitem[Greiner et al.(1994)]{gwh94} Greiner J., Wenzel W., Hudec R. \etal 
      (1994). {\it Gamma-Ray Bursts}, eds. G.J. Fishman \etal,
        AIP {\bf 307}, 408.


\bibitem[Greiner et al.(1995)]{gbk95} Greiner J., Bo\"{e}r M., Kahabka P., 
 Motch C., Voges W. (1995). NATO ASI {\bf C450}
{\it The Lives of the neutron stars}, eds. M.A. Alpar \etal,
Kluwer, p. 519.

\bibitem[Greiner et al.(1996)]{gbh96} Greiner J., Bade N., Hurley K., Kippen R.M., Laros J., (1996). 
3rd Huntsville workshop 1995, AIP \textbf{384}, p. 627.

\bibitem[Greiner et al.(2000)]{ghv00} Greiner, J., Hartmann D., Voges W., 
    \etal\ (2000). \textit{  A\&A} \textbf{353},  998.

\bibitem[Greiner et al.(2003)]{gkr03} Greiner J., Klose S., Reinsch K. \etal\ 
 (2003). \textit{ Nat.} \textbf{426}, 157.

\bibitem[Greiner et al.(2008)]{gbc08} Greiner J., Bornemann W., Clemens C. et al. (2008). \textit{
    PASP} \textbf{120}, 405.

\bibitem[Greiner et al.(2009a)]{gck09} Greiner J., Clemens C., Kr\"uhler T. 
  et al. (2009a). \textit{A\&A} \textbf{498}, 89.

\bibitem[Greiner et al.(2009b)]{gkf09} Greiner J., Kr\"uhler T., Fynbo J.P.U. 
  et al. (2009b). \textit{  ApJ}  \textbf{693}, 1610.

\bibitem[Greiner et al.(2010)]{gkk10} Greiner J., Kr\"uhler T., Klose S. 
   et al. (2010). \textit{ A\&A} (subm). 

\bibitem[Grindlay, Wright \& McCrosky(1974)]{gwm74} Grindlay J.E., Wright E.L.,
    McCrosky R.E.: (1974). \textit{ ApJ} {\bf 192}, L113.

\bibitem[Grupe et al.(2010)]{gbw10} Grupe D.,  Burrows D.N., Wu X.-F. et al.
  (2010). \textit{ ApJ} \textbf{711}, 1008. 

\bibitem[Grindlay(1999)]{grind99} Grindlay J.E. (1999). \textit{ ApJ} 510, 710.

\bibitem[Guetta \& Stella(2009)]{gul09} Guetta D. Stella L. (2009).
     \textit{A\&A} \textbf{498}, 329.

\bibitem[Heger, Langer \& Woosley(2000]{hlw00} Heger A., Langer N.,
     Woosley S.E. (2000). \textit{ApJ} \textbf{528}, 368.

\bibitem[Heise et al.(2001)]{heise01} Heise J., in't Zand J., Kippen M., 
 Woods P. (2001). in {\it GRBs in the afterglow Era}, Eds. E. Costa et al., 
 ESO-Springer, 16.

\bibitem[Harrison et al.(1995)]{hmp95} Harrison T.E., McNamara B.J., Pedersen H., \etal: (1995). \textit{
     A\&A} \textbf{297}, 465.

\bibitem[Hjorth et al.(1999)]{hjo99} Hjorth J., et al. (1999). \textit{
Sci.} \textbf{283}, 2073.

\bibitem[Hjorth et al.(2003)]{hjo03}  Hjorth, J., et al. (2003). \textit{
Nat.} \textbf{423}, 847. 

\bibitem[Hoyle(1946)]{hoy46} Hoyle F. (1946). \textit{MN} \textbf{106}, 343

\bibitem[Hudec et al.(1987)]{hbw87} Hudec R., Borovicka, J., Wenzel, \etal\ 
 (1987). \textit{  A\&A} {\bf 175}, 71.


\bibitem[Hurkett et al.(2008)]{hvo08} Hurkett C.P., Vaughan S., Osborne J., 
  \etal\ (2008). \textit{ ApJ} {\bf 679}, 587.

\bibitem[Hurley et al.(1999)]{hur95} Hurley K. (1999). \textit{  ApJS} \textbf{120}, 399.

\bibitem[Hurley et al.(1997)]{hur97} Hurley K., Costa E., Feroci M. et al. (1997). \textit{
  ApJ} \textbf{485}, L1. 

\bibitem[Inoue, Yamazaki \& Nakamura(2003)]{ino03} Inoue A.K., Yamazaki R., 
  Nakamura T. (2003). \textit{ ApJ} \textbf{601}, 644.

\bibitem[Iwamoto et al.(1998)]{iwa98} Iwamoto et al. (1998). \textit{ Nat.} \textbf{395}, 672.

\bibitem[Jelinek et al.(2006)]{jpk06} Jelinek M., Prouza M., Kubanek P. et al.
  (2006). \textit{ A\&A} \textbf{454}, L119.

\bibitem[Kaneko et al.(2007)]{krg07} Kaneko Y., Ramirez-Ruiz E., Granot J., Woosley S., et al. 
   (2007). \textit{ ApJ} 654, 385.

\bibitem[Kann et al.(2006)]{kkz06} Kann D.A., Klose S., Zeh A. (2006).
     \textit{ApJ} \textbf{641}, 993.

\bibitem[Kann et al.(2010)]{kkz10} Kann D.A., Klose S., Zhang B., et al.
   (2010). \textit{ApJ} (arXiv:0712.2186v2 from Sep. 2009).

\bibitem[Kehoe et al.(2002)]{kab02} Kehoe R., Akerlof C.W., Balsano R., et al. 
  (2002). \textit{ ApJ} 577, L159.

\bibitem[Kippen et al.(1994)]{kip94} Kippen R.M., \etal, (1994). {\it Gamma-Ray Bursts},
     eds. G.J. Fishman \etal, AIP {\bf 307}, 418.

\bibitem[Klose et al.(2003)]{khg03} Klose S., Henden A.A., Greiner J. et al. 
   (2003). \textit{ApJ} \textbf{592}, 1025.

\bibitem[Klotz et al.(2006)]{kgs06} Klotz A., Gendre B., Stratta G., et al.
 (2006). \textit{ A\&A} \textbf{451}, L39.

\bibitem[Klotz et al.(2009)]{kba09} Klotz, A., Bo\"er, M., Atteia, J.L., 
   Gendre, B. (2009). \textbf{AJ} \textbf{137},  4100.

\bibitem[Kobayashi(2000)]{kob00} Kobayashi S. (2000). \textit{ ApJ} \textbf{545}, 807.

\bibitem[Kogut et al.(2003)]{kogu03} Kogut A., Spergel D.N., Barnes C. et al.
  (2003). \textit{ ApJS} \textbf{148}, 161.
  
\bibitem[Kouveliotou et al.(1993)]{kou93} Kouveliotou, C., et al. (1993). 
  \textit{ApJ} \textbf{413}, L101. 

\bibitem[Krimm, Vanderspek \& Ricker(1994)]{kvr94} Krimm H., Vanderspek R.K., Ricker G.R.:
   (1994). {\it Gamma-Ray Bursts}, eds. G.J. Fishman \etal, AIP {\bf 307}, 423.

\bibitem[Krimm et al.(2007)]{kgm07} Krimm H.A., Granot J., Marshall F.E. 
   et al. (2007). \textit{ ApJ} \textbf{665}, 554.

\bibitem[Kr\"uhler et al.(2008)]{kkg08} Kr\"uhler T., K\"upc\"u Yolda\c{s} A.,
  Greiner J. et al. (2008). \textit{ ApJ} \textbf{685}, 376.

\bibitem[Kr\"uhler et al.(2009)]{kgm09} Kr\"uhler T., Greiner J., 
  McBreen S. et al. (2009). \textit{ ApJ} \textbf{697}, 758.

\bibitem[Kulkarni \& Rau(2006)]{kur06} Kulkarni S.R., Rau A. (2006). \textit{ 
    ApJ} \textbf{644}, L63.

\bibitem[Kumar \& Panaitescu(2000)]{kup00} Kumar P., Panaitescu A. (2000). \textit{
   ApJ} \textbf{541}, L51. 

\bibitem[K\"upc\"u Yolda\c{s} et al.(2006)]{aky06} K\"upc\"u Yolda\c{s} A., 
 Greiner J., Perna R. (2006). \textit{ A\&A} \textbf{457}, 115.

\bibitem[Lamb \& Reichart(2001)]{lar01} Lamb D.Q., Reichart D.E. (2001).   
in \textit{GRBs in the afterglow era}, eds. Costa \etal\, ESO-Springer, p. 226. 

\bibitem[Lamb, Donaghy \& Graziani(2005)]{ldg05} Lamb D.Q., Donaghy T.Q., 
 Graziani C. (2005). \textit{ ApJ} \textbf{620}, 355.

\bibitem[Lazzati, Ramirez-Ruiz \& Ghisellini(2001)]{lrg01} Lazzati D., 
   Ramirez-Ruiz E., Ghisellini G. (2001). \textit{ A\&A} \textbf{379}, L39. 

\bibitem[Lazzati et al.(2002)]{lrc02} Lazzati D., Rossi E., Covino S.,
Ghisellini G., Malesani D. (2002). \textit{ A\&A} 396, L5.

\bibitem[Ledoux et al.(2009)]{lvs09} Ledoux C., Vreeswijk P.M., Smette A. 
   et al. (2009). \textit{ A\&A} \textbf{506}, 661. 


\bibitem[Levesque et al.(2010)]{lbk10} Levesque E., Berger E., Kewley L., 
  Bagley M.M. (2010). \textit{AJ} \textbf{139}, 694.

\bibitem[Levinson et al.(2002)]{low02} Levinson A., Ofek E.O., Waxman E., 
  Gal-Yam A.  (2002). \textit{ ApJ} \textbf{576}, 923.


\bibitem[Li et al.(1996)]{li96} Li P., Hurley K., Sommer M., et al.
    (1993). \textit{ BAAS} \textbf{25}, 846.

\bibitem[Li et al.(2003)]{lfc03} Li W., Filippenko A.V., Chornock R., Jha S. 
   (2003). \textit{ ApJ} \textbf{586}, L9.

\bibitem[Lipkin et al.(2004)]{log04} Lipkin, Y.M., Ofek, E.O., Gal-Yam, A.
    et al. (2004). \textit{ApJ} \textbf{606}, 381.

\bibitem[Lithwick \& Sari(2001)]{lis01} Lithwick Y., Sari R. (2001). \textit{ 
   ApJ} \textbf{555}, 540.

\bibitem[Loeb \& Barkana(2001)]{loeb01} Loeb A., Barkana R. (2001). \textit{ 
    ARAA} \textbf{39}, 19. 

\bibitem[Madau, Della Valle \& Panagia(1998)]{mdp98} Madau P., Della Valle M.,
   Panagia N. (1998). \textit{MN} \textbf{297}, L17

\bibitem[Mao \& Mo(1998)]{mao98} Mao S., Mo H.J. (1998). \textit{ A\&A} \textbf{339}, L1. 

\bibitem[McBreen et al.(2010)]{mkr10} McBreen S., Kr\"uhler T., Rau A. 
  et al. (2010). \textit{ A\&A} \textbf{516} (in press; arXiv:1003.3885)

\bibitem[MacFadyen \& Woosley(1999)]{mfw99} MacFadyen A.I., Woosley S.E.
     (1999). \textit{ApJ} \textbf{524}, 262.

\bibitem[Mazzali et al.(2006)]{mdn06} Mazzali P.A., Deng J., Nomoto K. et al. 
  (2006). \textit{Nat.} \textbf{442}, 1018.

\bibitem[M\'{e}sz\'{a}ros, Rees \& Wijers(1998)]{mes98} M\'{e}sz\'{a}ros P., 
  Rees M.J., Wijers R.A.M.J. (1998). \textit{ ApJ} \textbf{499}, 301. 

\bibitem[M\'{e}sz\'{a}ros \& Rees(1999)]{mer99} M\'{e}sz\'{a}ros  P., Rees M.
  (1999). \textit{ MN} \textbf{306}, L39.

\bibitem[Metzger et al.(1997)]{mdk97} Metzger M.R., Djorgovski S.G., Kulkarni S.R. et al. (1997). \textit{ 
    Nat.} \textbf{387}, 878.

\bibitem[Metzger et al.(2008)]{mqt08} Metzger B.D., Quatert E., Thompson T.A.
     (2008). \textit{MN} \textbf{385}, 1455.

\bibitem[Miralda-Escud\'{e}(2003)]{mir03} Miralda-Escud\'{e} J. (2003). \textit{ 
   Sci.} \textbf{300}, 1904. 

\bibitem[Molinari et al.(2007)]{mvm07} Molinari E., Vergani S.D., Malesani D. et al. (2007). \textit{
  A\&A} \textbf{469}, L13.

\bibitem[Nakar \& Piran(2003)]{nap03} Nakar E., Piran T. (2003). \textit{ 
   New Astron.} \textbf{8}, 141.

\bibitem[Nishihara et al.(2003)]{nhk03} Nishihara E., 
   Hashimoto O., Kinugasa K. (2003). GCN 2118. 

\bibitem[Nomoto et al.(2004)]{nom04} Nomoto K., Maeda K., Mazzali P.A.
et al. (2004). in \textit{Stellar Collapse},
eds C.L. Fryer, \textit{ASSL} \textbf{302}, p. 277. 

\bibitem[Norris \& Bonnell(2006)]{nob06} Norris J.P., Bonnell J.T. (2006). 
   \textit{ApJ} \textbf{643}, 266.

\bibitem[Norris, Gehrels \& Scargle(2010)]{ngs10} Norris J.P., Gehrels, N., 
   Scargle, Jeffrey D. (2010). \textit{ApJ} \textbf{717}, 411.




\bibitem[Noterdaeme et al.(2008)]{nlp08} Noterdaeme P., Ledoux C., 
  Petitjean P., \& Srinand R. (2008). \textit{ A\&A} \textbf{481}, 327. 

\bibitem[Nousek et al.(2006)]{ncg06} Nousek J.A., Kouveliotou C., Grupe D., et al. (2006). \textit{
  ApJ} \textbf{642}, 389.

\bibitem[Oates et al.(2009)]{ops09} Oates S.R., Page M.J., Schady P. \etal\
  (2009). \textit{ MN} \textbf{395}, 490.

\bibitem[Oren, Nakar \& Piran(2005)]{onp05} Oren Y., Nakar E., Piran T. 
   (2005). \textit{ MN} \textbf{353}, L35.

\bibitem[Paczynski(1998)]{pac98} Paczynski B. (1998). \textit{ ApJ} \textbf{494}, L45.

\bibitem[Panaitescu \& Kumar(2001)]{pk01} Panaitescu A., Kumar, P. (2001). \textit{
  ApJ} \textbf{554}, 667.

\bibitem[Panaitescu \& Kumar(2002)]{pk02} Panaitescu A., Kumar, P. (2002). \textit{ 
   ApJ} \textbf{571}, 779.

\bibitem[Panaitescu \& Vestrand(2008)]{panaitescu08} Panaitescu A., Vestrand W.T. (2008). \textit{
  MN} \textbf{387}, 497. 
  
\bibitem[Paragi et al.(2010)]{paragi10} Paragi Z., Taylor G.B., 
   Kouveliotou C. et al. (2010). \textit{Nat.} \textit{463}, 516. 

\bibitem[Perley et al.(2008)]{pbb08} Perley D.A., Bloom J.S., Butler N.R. et al.
   (2008). \textit{ ApJ} \textbf{672}, 449. 

\bibitem[Perley et al.(2009)]{pcb09} Perley D.A., Cenko S.B., Bloom J.S.,
et al. (2009). \textit{ AJ} \textbf{138}, 1690.

\bibitem[Perna \& Loeb(1998)]{per98} Perna R., Loeb A. (1998). \textit{ ApJ} \textbf{509}, 
  L85. 

\bibitem[Perna, Raymond \& Loeb(2000)]{prl00}
   Perna, R., Raymond, J.,  Loeb, A. (2000). \textit{ ApJ} \textbf{533}, 658. 

\bibitem[Pihlstr\"{o}m et al.(2007)]{ptg07} Pihlstr\"{o}m, Y.M., Taylor, G.B., Granot, J., 
 Doeleman, S. (2007). \textit{ ApJ} \textbf{664}, 411. 

\bibitem[Piro(2001)]{piro01} Piro L. (2001). in {\it GRBs in the Afterglow era}, 
    eds. E. Costa, F. Frontera, J. Hjorth, Springer, Berlin, p. 97.

\bibitem[Piro \& Scarsi(2004)]{pis04} Piro L., Scarsi L. (2004). 
in {\it The Restless High-Energy Universe}, 
 May 2003, Amsterdam, Eds. E.P.J. van den Heuvel, R.A.M.J. Wijers, J.J.M. in 't Zand,
 \textit{Nucl. Phys. B} (Proc. Suppl). \textbf{132}, p. 3. 

\bibitem[Pizzichini et al.(1986)]{piz86} Pizzichini G., \etal\ (1986). 
 \textit{ApJ} {\bf 301}, 641.

\bibitem[Porciani, Viel \& Lilly(2007)]{pvl07} Porciani C., Viel M., Lilly S.J.
  (2007). \textit{ApJ} \textbf{659}, 218.

\bibitem[Price et al.(2003)]{pfk03} Price P.A., Fox D.W., Kulkarni S.R., et al. (2003). \textit{ Nat.} \textbf{423}, 844.

\bibitem[Podsiadlowski et al.(2004)]{pmn04} Podsiadlowski P., Mazzali P.A.,
    Nomoto K. et al. (2004). \textit{ApJ} \textbf{607}, L51.

\bibitem[Prochaska, Chen \& Bloom(2006)]{pcb06} Prochaska J.X., Chen H.-W.,
    Bloom J.S. (2006). \textit{ ApJ} \textbf{648}, 95. 

\bibitem[Prochter et al.(2006)]{pro06} Prochter G.E., et al. (2006). \textit{ 
    ApJ} \textbf{648}, L93. 

\bibitem[Quimby et al.(2006)]{qry06} Quimby R.M., Rykoff E.S., Yost S.A. \etal\
 (2006). \textit{ ApJ} \textbf{640}, 402.

\bibitem[Racusin et al.(2008)]{rks08} Racusin J.L., Karpov S.V., Sokolowski M.,
  et al. (2008). \textit{ Nat.} \textbf{455}, 183.

\bibitem[Racusin et al.(2009)]{rlb09} Racusin J.L., Liang E.W., Burrows D.N.
  \etal . (2009). \textit{ ApJ} \textbf{698}, 43.

\bibitem[Rau, Schwarz \& Greiner(2006)]{rgs06} Rau A., Greiner J., Schwarz R., 
  (2006). \textit{ A\&A} \textbf{449}, 79.

\bibitem[Rau et al.(2010)]{rsk10} Rau A., Savaglio S., Kr\"uhler T. et al.
  (2010). \textit{ ApJ} (subm.; arXiv:1004.3261).

\bibitem[Resmi et al.(2005)]{resmi+05} Resmi, L., Ishwara-Chandra, C.H., Castro-Tirado A.J. et al 
(2005). \textit{ A\&A} \textbf{440}, 477.

\bibitem[Rhoads(1997)]{rhoad97} Rhoads J.E., (1997). \textit{ ApJ} \textbf{487}, L1.

\bibitem[Rhoads(1999)]{rho99} Rhoads J.E., (1997). \textit{ ApJ} \textbf{525}, 737.

\bibitem[Richardson(2009)]{rich09} Richardson D. (2009). \textit{AJ} 
  \textbf{137}, 347.

\bibitem[Ricker et al.(2002)]{ric02} Ricker G.R., Atteia J.-L., Crew G.B. et al.
(2002). in \textit{GRB and Afterglow Astronomy 2001}
ed. G.R. Ricker \& R.K. Vanderspek (New York) AIP \textbf{662}, p. 3.

\bibitem[Rol et al.(2000)]{rol00} Rol, E., et al. (2000). \textit{
ApJ} \textbf{544}, 707. 

\bibitem[Romano et al.(2008)]{rcm08} Romano P., Campana S., Mignani R.P.,
   et al. (2008). {\it Vizier} Online Data Catalog 348:81221.

\bibitem[Rossi et al.(2010)]{rosk10} Rossi A., Schulze S., Klose S. et al. 
  (2010). \textit{A\&A} (subm.; arXiv:1007.0383)

\bibitem[Rykoff et al.(2004)]{rsp04} Rykoff E., Smith D.A., Price P.A. et al. 
  (2004). \textit{ ApJ} \textbf{601}, 1013.

\bibitem[Rykoff et al.(2005)]{raa05} Rykoff E., Aharonian F., Akerlof C.W., et al. (2005). \textit{
ApJ} \textbf{631}, 1032.

\bibitem[Rykoff et al.(2009)]{raa09} Rykoff E., Aharonian F., Akerlof C.W., et al. (2009). \textit{
     ApJ} \textbf{702}, 489. 

\bibitem[Sakamoto et al.(2008)]{shs08} Sakamoto T., Hullinger D., Sato G. 
  et al. (2008). \textit{ ApJ} \textbf{679}, 570.

\bibitem[Salvaterra et al.(2009)]{sdc09} Salvaterra, R., Della Valle, M., 
  Campana, S. et al. (2009). \textit{Nat.} \textbf{461}, 1258.

\bibitem[Sari, Piran \& Narayan(1998)]{spn98} Sari R., Piran T., Narayan R.
   (1998). \textit{ ApJ} \textbf{497}, L17. 

\bibitem[Sari(1999)]{sari99} Sari R. (1999). \textit{ ApJ} \textbf{524}, L43. 

\bibitem[Sari \& Piran(1999)]{sap99} Sari R., Piran T. (1999). \textit{ ApJ} \textbf{517}, L109.

\bibitem[Savaglio et al.(2009)]{sgb09} Savaglio, S., Glazebrook, K., Le Borgne, D. (2009). \textit{ ApJ}
   \textbf{691}, 182.


\bibitem[Schaefer et al.(1984)]{sbb84} Schaefer B.E., Bradt, H., Barat, C. 
  \etal . (1984). \textit{ ApJ}  {\bf 286}, L1.

\bibitem[Schady et al.(2007)]{smp07} Schady P., Mason K.O., Page M.J. et al.
    (2007). \textit{MN} \textbf{377}, 273.

\bibitem[Schady et al.(2010)]{spo10} Schady P., Page M.J., Oates S.R. et al.
    (2010). \textit{MN} \textbf{401}, 2773.

\bibitem[Schmidt(2000)]{msc00} Schmidt M. (2000). \textit{ ApJ} \textbf{552}, 36. 

\bibitem[Smartt et al.(2002)]{svr02} Smartt S.J., Vreeswijk P.M.,
   Ramirez-Ruiz E. et al. (2002). \textit{ApJ} \textbf{572}, L147.

\bibitem[Soderberg et al.(2010)]{scp10} Soderberg A.M., Chakraborti S., Pignata G. et al. (2010). \textit{
  Nat.} \textbf{463}, 513.

\bibitem[Spruit(2002)]{spr02} Spruit H. (2002). \textit{A\&A} \textbf{381}, 923.

\bibitem[Stanek et al.(2003)]{sta03} Stanek K., et al. (2003). \textit{ 
  ApJ} \textbf{591}, L17.

\bibitem[Steele et al.(2010)]{sms10} Steele I.A., Mundell C.G., Smith R.J., Kobayashi S., 
  Guidorzi C. (2010). \textit{ Nat.} \textbf{462}, 767.

\bibitem[Stratta et al.(2004)]{sfa04} Stratta G., Fiore F., Antonelli L.A.
   et al. (2004). \textit{ApJ} \textbf{608}, 846.

\bibitem[Sudilovsky et al.(2007)]{ssv07} Sudilovsky V., Savaglio S., 
   Vreeswijk P.M. et al. (2007). \textit{ ApJ} \textbf{669}, 741.

\bibitem[Tagliaferri et al.(2005)]{tgc05} Tagliaferri, G., Goad, M., 
    Chincarini, G. et al. (2005). \textit{Nat.} \textbf{436}, 985.

\bibitem[Tanvir et al.(2009)]{tfl09} Tanvir N.R., Fox D.B., Levan A.J. 
et al. (2009). \textit{ Nat.} \textbf{461}, 1254.

\bibitem[Taylor et al.(2004)]{tfb04} Taylor, G.B., Frail, D.A., Berger, E., Kulkarni, S.R. (2004). \textit{
 ApJ} \textbf{609}, L1. 

\bibitem[Tejos et al.(2009)]{tlp09} Tejos N., Lopez S., Prochaska J.X. et al.
   (2009). \textit{ ApJ} \textbf{706}, 1309. 

\bibitem[Troja et al.(2008)]{tko08} Troja E., King A.R., O'Brien P.T.
      et al. (2008). \textit{MN} \textbf{385}, L10.

\bibitem[Vanden Berk et al.(2002)]{vlw02} Vanden Berk D.E., Lee B.C., Wilhite B.C. et al. (2002). \textit{
  ApJ} \textbf{576}, 673


\bibitem[Van der Horst et al.(2008)]{horst07} Van der Horst, A.J., Kamble, A., 
  Resmi, L. et al. (2008). \textit{ A\&A} \textbf{480}, 35.

\bibitem[Vanderspek, Krimm \& Ricker(1994)]{vkr94} Vanderspek R., Krimm H.A., Ricker G.R.:
    (1994). {\it Gamma-Ray Bursts}, eds. G.J. Fishman \etal, AIP {\bf 307}, 438.

\bibitem[Vanderspek, Krimm \& Ricker(1995)]{vkr95} Vanderspek R.K., Krimm H.A.,
 Ricker G.R.: (1995). \textit{
Ap\&SS} \textbf{231}, 259.

\bibitem[Van Paradijs et al.(1997)]{vp97} Van Paradijs, J. Groot, P.J., 
  Galama, T. \etal\ (1997). \textit{Nat.} \textbf{386}, 686. 

\bibitem[Vergani et al.(2009)]{vpl09} Vergani S.D., Petitjean P., Ledoux C.
et al. (2009). \textit{ A\&A} \textbf{503}, 771.

\bibitem[Vestrand, Borozdin \& Casperson(2004)]{vbc04} Vestrand W.T., 
   Borozdin K., Casperson D.J. (2004). \textit{ AN} \textbf{325}, 549.

\bibitem[Vestrand et al.(2005)]{vest05} Vestrand W.T., Wozniak, P.R., 
  Wren, J.A. et al. (2005). \textit{ Nat.} \textbf{435}, 178.

\bibitem[Vestrand et al.(2006)]{vest06} Vestrand W.T., Wren, J.A., 
  Wozniak, P.R. et al. (2006). \textit{ Nat.} \textbf{442}, 172.

\bibitem[Villasenor et al.(2004)]{villa04} Villasenor J., Ricker G., Vanderspek R.
et al. (2004). in \textit{Proc. 35th COSPAR Sci. Assembly}, July 2004, Paris, p. 1225.


\bibitem[Villasenor et al.(2005)]{vlr05} Villasenor J., Lamb D.Q., Ricker G. et al (2005). \textit{
Nat.} \textbf{437}, 855. 

\bibitem[Vink \& de Koter(2005)]{vdk05} Vink J.S., de Koter A. (2005).
     \textit{A\&A} \textbf{442}, 587.

\bibitem[Virgili et al.(2010)]{vzo10} Virgili F.J., Zhang B., O'Brien P.T.,
   Troja E. (2010). \textit{ApJ} (subm.; arXiv:0909.1850)

\bibitem[Vreeswijk et al.(2004)]{vel04} Vreeswijk P.M., Ellison S.L., 
   Ledoux C. et al. (2004). \textit{ A\&A} \textbf{419}, 927. 

\bibitem[Vreeswijk et al.(2007)]{vls07} Vreeswijk P.M., Ledoux C., 
    Smette A. et al. (2007). \textit{ A\&A} \textbf{468}, 83. 

\bibitem[Waxman(1997)]{wax97} Waxman E. (1997). \textit{ ApJ} \textbf{489}, L33. 

\bibitem[West et al.(2008)]{wmb08} West J.P., McLin K., Brennan T. et al.
  (2008).  GCN 8617. 

\bibitem[Wijers \& Galama(1999)]{wig99} Wijers, R.A.M.J., Galama T.J. (1999). \textit{ 
 ApJ} \textbf{523}, 177.

\bibitem[Wijers et al.(1999)]{wij99} Wijers, R.A.M.J., et al. (1999). \textit{
ApJ} \textbf{523}, L33. 

\bibitem[Woosley, Heger \& Weaver(2002)]{whw02} Woosley S.E., Heger A.
     Weaver T.A. (2002). \textit{Rev. Mod. Phys.} \textbf{74}, 1015.

\bibitem[Woosley \& Bloom(2006)]{wob06} Woosley S.E., Bloom J.S. (2006). \textit{ARAA} \textbf{44}, 507.

\bibitem[Yoon \& Langer(2005)]{yol05} Yoon S.-C., Langer N. (2005).
     \textit{A\&A} \textbf{443}, 643.

\bibitem[Yost et al.(2002)]{yfh02} Yost S.A., Frail D.A., Harrison F.A. et al.
  (2002). \textit{ ApJ} \textbf{577}, 155.

\bibitem[Yost et al.(2006)]{yar06} Yost S.A., Alatalo K., Rykoff E.S., et al. 
  (2006). \textit{ ApJ} \textbf{636}, 959.

\bibitem[Yuan \& Rujopakarn(2008)]{yur08} Yuan F., Rujopakarn W. (2008). GCN 8536.

\bibitem[Zhang et al.(2006)]{zfd06} Zhang B., Fan Y.Z., Dyks J. et al. (2006). \textit{
ApJ} \textbf{642}, 354.

\bibitem[Zhang et al.(2009)]{zzv09} Zhang, B., Zhang, B.-B., Virgili, F.J.
   et al. (2009). \textit{ApJ} \textbf{703}, 1696.

\end{thereferences}
\end{document}